# ELECTRON DIFFRACTION BASED TECHNIQUES IN SCANNING ELECTRON MICROSCOPY OF BULK MATERIALS

(Diffraction based techniques in SEM of Bulk Materials)

Angus J. Wilkinson and Peter B. Hirsch

Department of Materials, University of Oxford, Parks Road, Oxford, OX1 3PH , U.K.

# DIFFRACTION BASED TECHNIQUES IN SCANNING ELECTRON MICROSCOPY OF BULK MATERIALS

**Angus J. Wilkinson and Peter B. Hirsch**

**Department of Materials, University of Oxford, Parks Road, Oxford, OX1  3PH , U.K.**

**Abstract**

The three scanning electron microscope diffraction based techniques of electron channelling patterns (ECPs), electron channelling contrast imaging (ECCI), and electron back scatter diffraction (EBSD) are reviewed. The dynamical diffraction theory is used to describe the physics of electron channelling, and hence the contrast observed in ECPs (and EBSD) and ECCI images of dislocations.  Models for calculating channelling contrast are described and their limitations discussed.  The practicalities of the experimental methods, including detector-specimen configurations, spatial resolution and sensitivities are given.  Examples are given of the use of ECCI for imaging and characterising lattice defects, both individually and in groups, in semiconductor heterostructures and fatigued metals.  Applications of the EBSD technique to orientation determination, phase identification and strain measurement are given and compared with use of ECPs..  It is concluded that these techniques make the SEM a powerful instrument for characterising the local crystallography of bulk materials at the mesoscopic scale.

# CONTENTS

# I. INTRODUCTION

In 1967 Coates (1967) reported on the strong crystal orientation dependence of the intensity of high energy (or primary) back scattered electrons (BSEs) which he demonstrated by imaging large areas of bulk semiconductor single crystals at low (times 10) magnifications in the scanning electron microscope (SEM). In the paper immediately following this Booker, Shaw, Whelan and Hirsch (1967) confirmed Coates' observations and provided a theoretical interpretation of the phenomenon. This theory was given in terms of the Bloch wave description of the dynamical diffraction theory that Howie and Whelan (1961, 1962) had developed to describe diffraction effects in electrons transmitted through thin foils. Electron channelling patterns (ECPs) are now a feature available on many modern SEMs. In this early paper Booker, Shaw, Whelan and Hirsch (1967) also suggested that since close to the Bragg condition the BSE intensity varies rapidly with orientation it could be possible to use the effect to image crystal defects near the surface of a suitably oriented bulk specimen in the SEM. This forms the basis of the technique now called electron channelling contrast imaging (ECCI), which received attention throughout the 1970s and has been further developed over the last 6 years at Oxford.

Another electron diffraction technique capable of giving local crystallographic information from bulk specimens was described by Alam, Blackman and Pashley (1954) in the 1950s, though it was not applied in the SEM until the work of Venables and Harland (1973) . The technique has been referred to by several different names during its development but electron back scatter diffraction (EBSD) now appears to be emerging as the dominant one. In the EBSD method a stationary electron probe is focused onto a small region of a crystal and the angular distribution of BSEs emerging from the specimen is recorded. The form of the Kikuchi bands seen in the EBSD patterns is similar to that of channelling bands in ECPs, as a result of the reciprocity theory which connects the two types of pattern.

This paper sets out to review the progress made both in understanding SEM based electron diffraction phenomena, and in using them to generate information of importance to material scientists. Section 2 describes the basic channelling mechanism and the models that have been developed to simulate both ECPs and ECCI, and ends with a brief outline of applications and limitations of ECPs. The development of ECCI for imaging and characterising lattice defects is discussed in section 3, along with more recent work on analysing longer range strain fields. The EBSD technique is reviewed in section 4, with emphasis on the quantitative information that it can provide. Finally, in section 5, we draw the conclusion that in combination these SEM based diffraction techniques are of great importance to materials scientists.

## II. THE ELECTRON CHANNELLING MECHANISM AND ELECTRON CHANNELLING PATTERNS

### II A  Experimental Formation of Electron Channelling Patterns

An ECP, such as the example in fig. (1), records the changes in the BSE (or secondary electron) intensity that occur as the beam is rocked through a range of incident angles relative to the crystal planes. This was originally achieved simply by observing a large flat single crystal at low magnification, so that the double deflection scanning system (fig. 2a) not only varies the beam position but also alters its direction. Fig. (2a) makes it clear that beam moves a considerable distance over the specimen to enable the beam direction to be altered by a reasonable amount. With the excitation of the second scan coils reduced then the ray bundle can be deflected back through the optic axis at some position below the final aperture, and if the specimen were moved to this position then beam rocking would be achieved without the beam moving large distances across the specimen (Van Essen and Schulson 1969). In practice such a method requires the use a large final aperture to allow the beam to be rocked through a reasonably large range of angles, and the divergence then needs to be controlled with an aperture further up the column. With such a double deflection method ECPs can be obtained from areas down to ~50 µm in diameter for a scan angle of ~±5° (Van Essen and Schulson 1969).

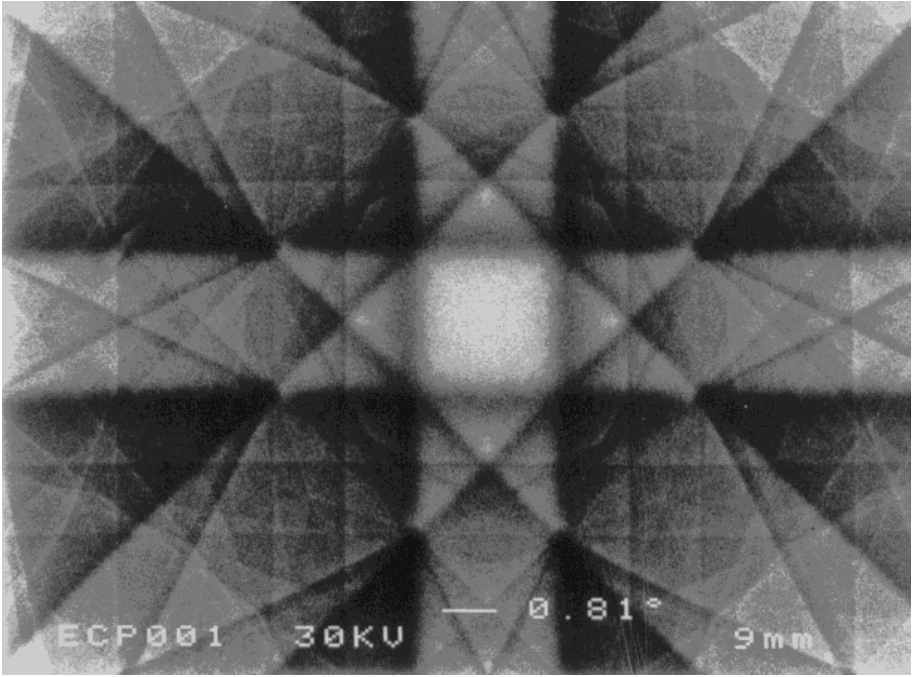

*Fig. 1      An example ECP obtained from Si specimen at 30 keV, showing a <100> zone axis.*

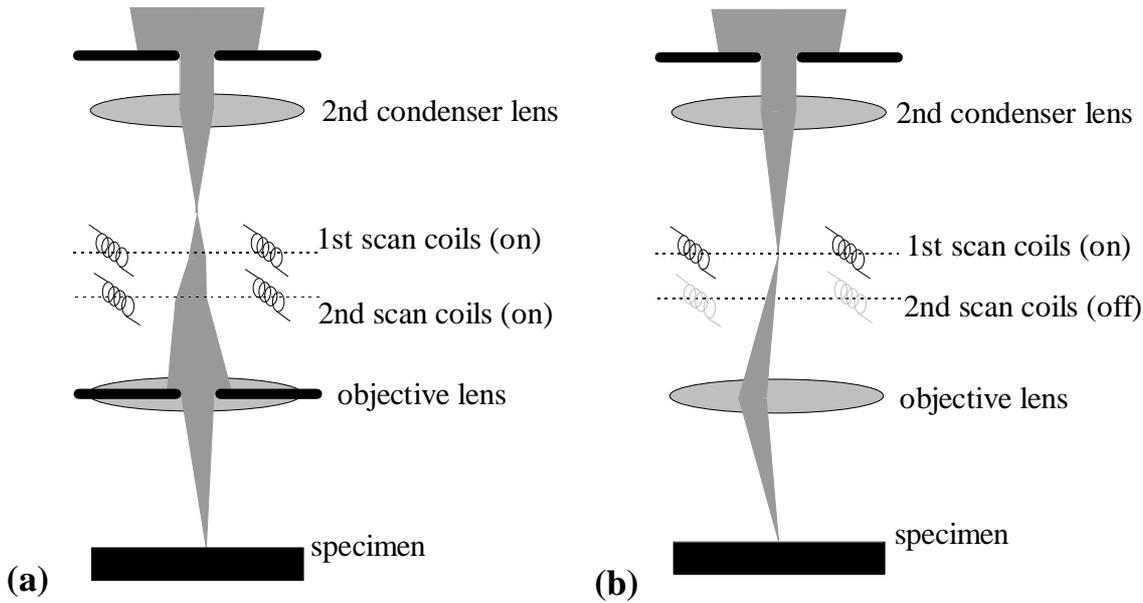

*Fig. 2      Ray diagrams showing (a) double deflection method of beam scanning for surface imaging and (b) deflection focus method of beam rocking*

In most SEMs the deflection focusing method (fig. 2b) has been adopted for generating the ECP. The final condenser lens is focused on the back focal plane of the objective and the upper scan coils are used to deflect the ray bundle across the objective lens. The focusing action of the objective lens then acts to bend the ray bundle back through the focal point on the optic axis which is set-up to be on the specimen surface. The maximum extent to which the beam can be rocked is determined by the size and position of the objective aperture and is typically limited to 10°-15°. Also the imperfection in particular the spherical aberration of the magnetic lenses in SEMs means that the beam is not truly rocked about a point. The spherical aberration of the objective lens results in the more highly inclined rays being brought back to the optic axis at a shorter distance than those that are less inclined. The resulting minimum disc of confusion $d_{ecp}$ sets the resolution of the ECP and is given by $d_{ecp} = 1/2\, C_s\, \theta_m^3$ and depends strongly on the maximum rocking angle ($\theta_m$) selected and the spherical aberration coefficient ($C_s$) of the objective lens. Booker and Stickler (1972) attained a resolution of 3 µm at a total rocking angle ($2\theta_m$) of 9°, by using a short working distance (1 mm) to minimise $C_s$; at a more usual working distance (6 mm) $d_{ecp}$ had increased to 20 µm for the same rocking angles.

Although the ray bundle is scanned through a large range of angles, at any one time it only has a narrow divergence, so that if the excitation of the final lens were reduced as the ray bundle was more steeply inclined then the effects of spherical aberration could be corrected for. This idea was implemented by Van Essen (1971) who adopted a spiral scan path, in preference to the usual Cartesian scan, so that the corrections to the current supplied to the highly inductive objective lens could be made more gradually. Van Essen (1971) reported that these dynamic corrections allowed a resolution of ~2 µm to be achieved.

The best resolution of the ECP technique routinely obtainable on modern SEMs is approximately 2 µm, though on many instruments the resolution is markedly worse than this, often by an order of magnitude or more.

## II B  The Channelling Mechanism

Booker, Shaw, Whelan and Hirsch (1967) gave a clear qualitative account of the theory underlying electron channelling, in the paper immediately following that reporting Coates' (1967) original observations of 'Kikuchi-like reflection patterns'. In their model the fast incident electrons are described inside the crystal by a superposition of many Bloch wave functions. The extent to which each of these waves is excited depends on the incident beam direction, due to the boundary conditions imposed by the need to maintain continuity of the wave function and its derivative across the free surface. Here we shall consider the situation when the incident beam is close to the Bragg condition for a given set of lattice planes so that only two Bloch waves dominate, though the ideas extend readily to situations where many such waves are excited. The two types of Bloch wave are depicted in fig. (3). The first of the Bloch waves has the maximum likelihood of finding an electron centred on lines of atoms running through the crystal. The time averaged probability density is high close to the atomic sites and interactions with the atomic nuclei and inner shell electrons are encouraged, resulting in relatively strong back scattering and heavy attenuation of the wave. Conversely the type 2 wave has minima centred on the atomic sites, while the high probability density is located in the channels between the lattice planes. For electrons in this wave the chances for scattering are much reduced, and the wave is transmitted through the crystal with less rapid attenuation. As previously mentioned the initial excitation of the two Bloch waves depends on the incident beam direction, or in other words on the deviation from the Bragg condition. Exactly at the Bragg angle the two waves are equally excited. At larger angles (see fig. 4) or positive deviations from the Bragg condition the type two wave is preferentially excited; this leads to the 'anomalous' transmission seen in thick foils, and in bulk crystals leads to penetration to enhanced depths. This results in a low yield of BSEs due firstly to a low probability of any initial large angle (or back) scattering from atomic nuclei, and secondly to the longer average distance that a multiply scattered electron would need to traverse in order to escape back to the crystal surface. When the incident beam is at an angle less than the Bragg angle to the diffracting planes (negative deviation from the Bragg condition) the type 1 Bloch wave dominates (see fig. 4), which in thin foils leads to poor transmission through the crystal. For a bulk crystal such conditions cause the BSE intensity to be increased, since the chances of an initial large angle scattering event is increased, and also the average penetration into the crystal is reduced so that multiply scattered electrons are more likely to escape the specimen.

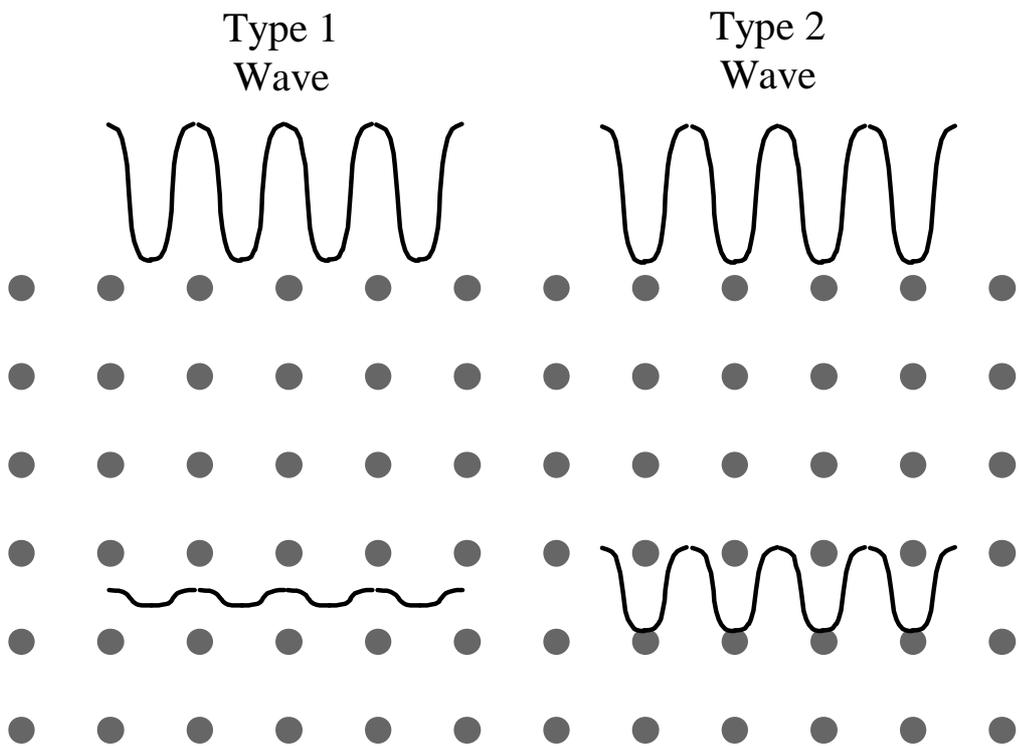

*Fig. 3    Bloch wave model of channelling*

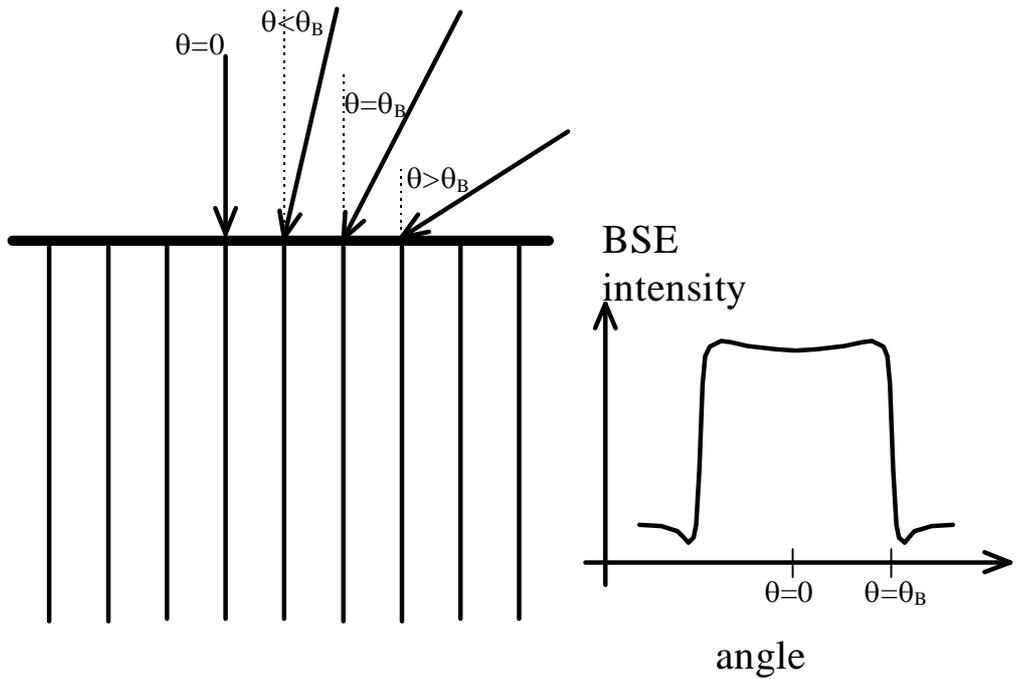

*Fig. 4    Schematic diagram of the form of channelling band contrast in relation to deviation from the Bragg condition.*

### II C  Models of Electron Channelling

We see from the above description that a complete model must incorporate both the initial dynamical diffraction aspect of the problem and the subsequent multiple scattering of the electrons that finally leads to some being emitted from the specimen as BSEs.  A great deal has already been written about the dynamical diffraction of electrons and the reader is referred to one of the many texts on diffraction contrast in the transmission electron microscope (e.g. Hirsch, Howie, Nicholson, Pashley and Whelan 1978).  It is in the treatments of the second aspect of the electron back scattering problem that the proposed models show the greatest difference.   Broadly the models can be split into two types, namely single scattering models and multiple scattering models.

*(a) single scattering models*

The general framework assumed within the single scattering models is shown schematically in fig. (5).  The electron intensity remaining in the initially excited Bloch waves states is denoted $I_E$ and its decrease with depth into the crystal is described by the dynamical diffraction theory.  Electrons are removed from these Bloch wave states (absorbed in TEM terms) either by scattering through a large ($\geq \pi/2$) angle into a backward travelling plane wave with intensity $I_B(z)$ or through smaller angles into a forward travelling plane wave, with intensity denoted $I_F(z)$. The only further scattering events that are considered are large angle scattering of electrons from the forward travelling into the backward travelling plane waves. Contributions to $I_B(z)$ thus occur through the action of a single large angle scattering event, and multiple scattering is ignored. Back scattering coefficients $p^{(j)}$ are defined as the fraction of intensity scattered from Bloch wave j through an angle greater than $\pi/2$, while a similar coefficient $p^{(0)}$ describes the probability of back scattering from the forward to backward travelling plane waves. The BSE intensity, given by $I_B(0)$, is then found by integrating an equation written for the change $dI_B(z)$ in the intensity of the backward travelling plane wave occurring across a small slice of the crystal. In Clarke and Howie's (1971) model $I_B(z)$ is assumed to remain small resulting in a kinematic approximation for which $I_F(z)$ is given simply by $1-I_E(z)$. This kinematic approximation breaks down for large specimen thickness, since eventually $I_E$ falls to zero as electrons are removed from the Bloch wave states, while $I_F$ tends to 1 at large depths. Backscattering from the forward travelling plane wave can thus occur at all large depths and contribute to $I_B(z)$ with the resulting in $I_B(0) \rightarrow \infty$ for an infinitely thick specimen. This problem was avoided by Clarke and Howie (1971) by terminating the integration of contributions to $I_B(0)$ at some maximum depth $t_m$, at which the intensity remaining in the initial Bloch wave states had been reduced to a fixed small value (in fact $10^{-6}$ was used). This allowed the variations in BSE intensity to be calculated but not the absolute values required to make predictions of achievable signal and contrast levels. Marthinsen and Høier (1988) assumed that the only events leading to the escape of a BSE is backscattering directly from the Bloch waves, and that no scattering from $I_F$ to $I_B$ is possible (i.e. $p^{(0)} = 0$). This completely removes the kinematic term from their model.

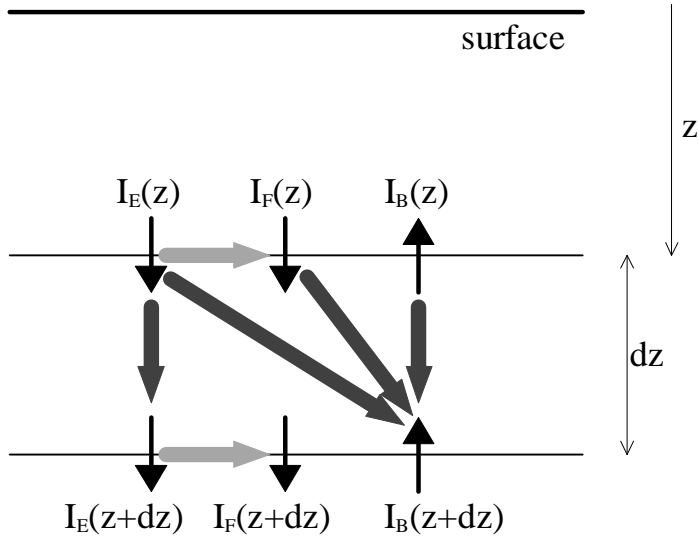

*Fig. 5*    *Scattering scheme assumed in the single scattering models. $I_E$, $I_F$ and $I_B$ denote electrons in the initial Bloch waves, forward scattered electrons and backscattered electrons respectively. No scattering from $I_B$ into $I_F$ is allowed, and $I_F$ is obtained from a kinematic treatment.*

In their calculations Clarke and Howie (1971) assumed the backscattering probability to be simple delta functions at each atomic site, with the result that the $p^{(j)}$s were independent of j. Subsequent work by Reimer, Heilers and Saliger (1986), Marthinsen and Høier (1988), and recently Rossouw, Miller, Josefsson and Allen (1994) have improved on this last point by calculating $p^{(j)}$ from models of phonon or thermal diffuse scattering (TDS) which is the dominant cause of the large angle scattering process. Use of the Rutherford scattering cross-sections (Marthinsen and Høier 1988 and Rossouw, Miller, Josefsson and Allen 1994) within the TDS models leads to convenient analytical expressions for $p^{(j)}$, with integration either being undertaken over the $2\pi$ steradian solid angle of backscattering, or limited to the angular range subtended by the BSE detector used experimentally as in the work of Rossouw *et al* (1994). Reimer *et al* (1986) have incorporated Mott cross-sections into their calculations through the use of tabulated numerical data. Although Rutherford cross-sections can show significant discrepancies from the more correct Mott cross-sections, especially for high atomic numbers and low beam energies, their use is perhaps unnecessary given the coarse approximation made in neglecting multiple scattering.

As an example fig. (6) compares results from the many (75) beam calculations of Rossouw *et al* (1994) with an experimental ECP of a <111> zone axis in a thin foil of Si obtained with BSEs using an incident beam energy of 300 keV. The single scattering approximation is more appropriate for such thin foil situations (which arise when examining the effects of channelling on x-ray emission); however for bulk specimens multiple scattering must be considered.

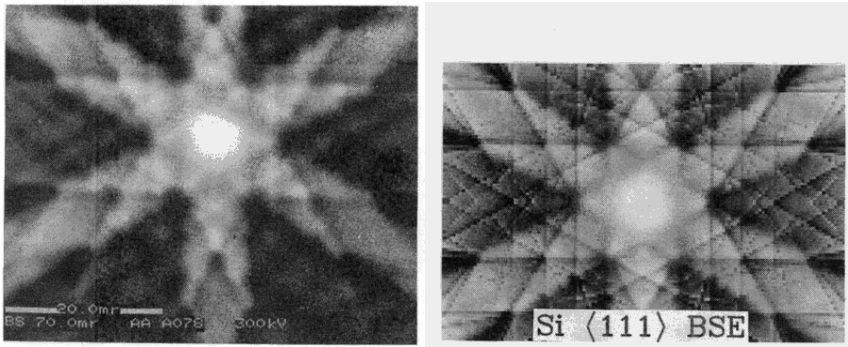

*Fig. 6    Comparison of (a) experimental, and (b) simulated ECP from a thin (0.4 µm) Si foil viewed along <111> with 300 keV electrons.  (from Rossouw et al 1994)*

*(b) multiple scattering models*

The simplest treatment in which some account of multiple scattering is included is the so called forward-backward approximation that was devised by Hirsch and Humphreys (1970). The scattering scheme imposed is shown diagramatically in fig. (7). Similarities with the single scattering model are evident (compare figs. 5 and 7), and again electrons are removed from the initial Bloch waves into two plane waves, one forward and one backward propagating. However, in order to make some allowance for multiple scattering across each slab of material a fraction of the electrons intensity $I_F$ is backscattered into $I_B$, while a similar fraction of electrons in $I_B$ also reverses direction. $I_F$ is no longer given by the simple kinematic expression of the single scattering models. Assuming that the electron intensity is conserved (i.e. $I_E+I_F+I_B=1$) differential equations can be written and integrated to find $I_B(0)$. For perfect crystals the integration can be done analytically while for crystals containing dislocations or other strain fields numerical methods must be used. The forward-backward approximation takes no account of energy losses and as a result also has problems for specimens of large thickness. In fact within this model the electrons effectively have an infinite range with the result that $I_B(0) \rightarrow 1$ for an infinitely thick specimen. Spencer, Humphreys and Hirsch (1972) solved this problem by imposing a maximum effective thickness $t_b$ for bulk specimens, which reflects the maximum depth to which electrons can penetrate into the crystal and still escape from the surface after backscattering. For values of $t_b$ taken as a fixed fraction (0.4) of the Bethe range the calculated average yields of BSEs are typically between 40 and 60% of those determined experimentally. Although the shapes of calculated rocking curves are in broad agreement with experiment the calculated contrast levels tend to be somewhat too small. A more elaborate method was used by Sandström, Spencer and Humphreys (1974) in which energy losses were introduced. $I_E$, $I_F$, and $I_B$ then become a set of functions at discrete energy levels and scattering to lower energies was treated through a quasi-continuous implementation of the Bethe energy loss law. Despite the more involved computations the calculated channelling contrast and the average BSE yields remained little changed from that achieved with the previous theory of Spencer *et al* (1972).

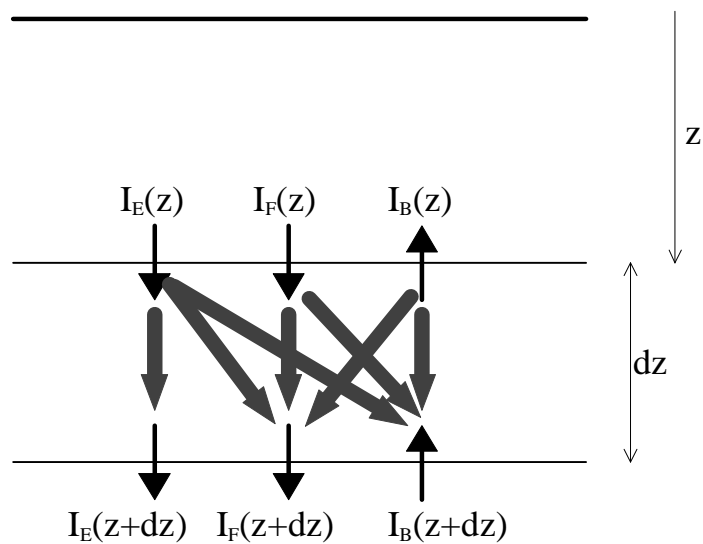

*Fig. 7    Scattering scheme assumed in the forward-backward scattering models.  $I_E$, $I_F$ and $I_B$ denote electrons in the initial Bloch waves, forward scattered electrons and backscattered electrons respectively.  Multiple scattering between $I_B$ and $I_F$ is allowed.*

The quantitative predictions of calculations within the forward-backward approximation are of limited success due to the neglect of medium angle (10° to 90°) scattering events. The low BSE yields of the forward-backward models are due to the neglect of the possibility of an electron reversing direction through the cumulative action of medium angle scattering events. Furthermore, as with large angle scattering, the scattering of the electrons through medium angles requires interaction with the massive atomic nuclei and so is dominated by phonon scattering. Thus whenever there is an increased probability of electrons in the initial Bloch wave states being scattering directly through >90°, there is also an increased chance of a medium angle event after which further interactions can lead to backscatter. Thus for incident beam directions giving a high BSE yield in the forward-backward approximation, the yield will be further increased by the action of medium angle scattering and hence lead to larger channelling contrast levels.

Spencer & Humphreys (1980) used a transport equation approach in an attempt to incorporate the effects of medium angle scattering. They found a solution for the axisymmetric case at normal incidence, and for thin specimens the thickness variation of the channelling contrast and background BSE level were in better agreement with experiment than calculated with the forward-backward approximation. However, energy loss effects were ignored so that for thicker specimens an effective thickness $t_b$ again had to be employed, though in this work it was empirically set so that the mean BSE yield was in agreement with experiment.

In an alternative approach, dynamical diffraction theory has been used to simulate the initial scattering event, and results from Monte-Carlo models used to provide the probabilities of the electrons being emitted from the surface at a given angle, and with a given energy (Yamamoto, Mori and Ishida 1978). Since the channelling contrast is low the computer intensive Monte-Carlo simulations need to be made for many possible electron trajectories in order to remove artefacts due to statistical fluctuations, and new simulations are required for each new material, or incident beam energy. Simulations of channelling bands showed that the contrast was asymetric for tilted specimens, when not under symmetric Laue diffraction conditions. The contrast was also found to be dependent on the range of takeoff angles assumed for the detector, an effect that cannot be studied within the forward-backward approximation.

The most recent advances have been made by Dudarev, Rez and Whelan (1995). Scattering of electrons from the initial Bloch waves forms a source function in an inhomogeneous transport equation describing the subsequent multiple scattering. Dudarev *et al* have developed numerical methods for solving this transport equation based on algorithms given by Fathers and Rez (1982). The model has been used to calculate the effect of temperature on the average BSE yields from polycrystals of different atomic number (Dudarev and Whelan 1994), and correctly simulates the reversal of atomic number contrast that occurs at low incident beam energies (Dudarev : private communication). Many beam calculations of ECPs are in good agreement with experiment. Energy loss processes are fully incorporated into the model and calculated energy spectra from dark regions of ECP show a marked decrease at the high energy peak, with a slight increase along the low energy tail compared to spectra from the bright channelling bands, which is in agreement with experimental observation of Wolf, Coane and Everhart (1970). Contrast reversals seen with the important tilted specimen geometry are accurately mimicked in the simulations. Fig. (8) shows two simulated ECP from a Si (001) specimen in which the incident beam is tilted through 55° (top of the pattern) to 75° (bottom of the pattern) from normal incidence. The dramatic difference between the two patterns in fig. (8) is caused by the different positions assumed for the detectors: in (a) low takeoff angle electrons are used while in (b) high takeoff angle electrons that would in practice be captured when a detector under the pole piece is used. The contrast reversal shown in (b) is observed experimentally (Ichinokawa, Nishimura and Wada 1974), and the details are found to depend not only on the detector's position but also on its energy response. Furthermore, Dudarev, Rez and Whelan (1995) have made a perturbation expansion of their equations, the use of which reduces the computational requirements, with which it is possible to calculate ECCI contrast from crystal defects.

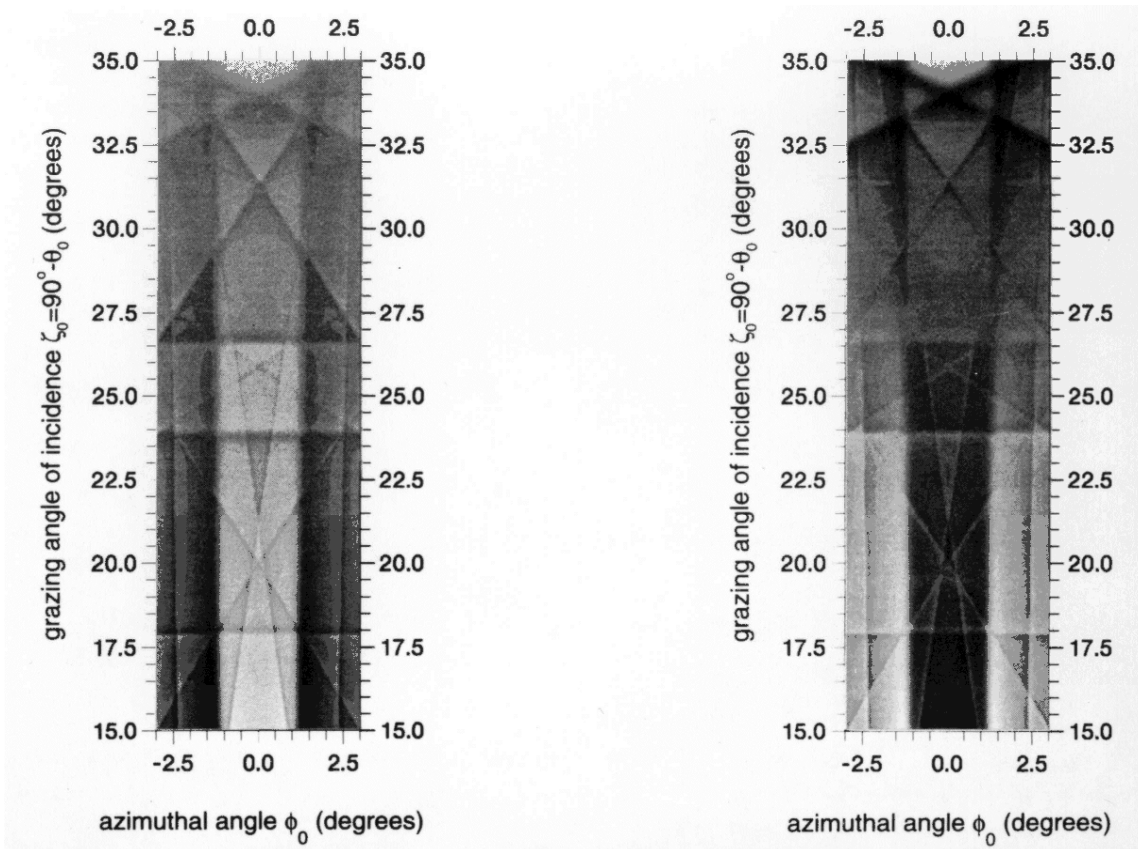

*Fig. 8    Comparison of simulated ECP for a tilted Si specimen showing a {220} channelling band with the beam incident at varying glancing angles of 15° to 35° from the surface plane of the specimen. (a) for a low takeoff angle detector in the forward scattering direction 10° away from the surface tangent, and (b) for a high takeoff angle detector 140° away from the surface tangent taken in the forward scattering direction. (from Dudarev et al 1995*

### II D  Applications of ECPs

We shall give here only brief examples of the uses to which ECPs have been put and for further examples refer the reader to several reviews of electron channelling which have already been made (Schulson 1977, Joy, Newbury and Davidson 1982, and Davidson 1984).

The orientation of crystals, and crystal facets can be obtained from ECPs.  Davidson's (1974) measurements of fracture plane orientations in Mo 15 atom % Re crystals made use of a comparison technique, whereby patterns obtained from the area of interest are visually compared to a previously produced montage of ECPs covering all possible crystal orientations.  Alternatively, if the position of a known zone axis and the inclination of a known band passing through it can be identified then the ECP can be used to establish the full grain orientation directly, as for example in the work of Ayers and Joy (1972).  This type of analysis is often hampered by the limited capture angle obtainable in a ECP which is typically limited to 10°-15°. Typically the orientation can be determined to an accuracy of ±0.5°.

The effects of lattice imperfections are readily apparent in ECP, which become increasingly diffuse as more defects are introduced into the crystal. Stickler, Hughes and Booker (1971) found that the angular width of the finest resolvable lines increased with dislocation density for a range of materials. Ruff (1976) used the reduction in contrast of a low order channelling band, which he correlated with plastic strain in a calibration specimen, to allow subsequent measurements of plastic strain distributions around wear tracks in iron. Davidson and Booker (1970) showed that ion implantation damage also degrades the sharpness of ECP and more recently Page, McHargue and White (1991) have looked at this effect again and used it to show the shallow (<150 nm in sapphire at 35 keV) information depth of electron channelling compared to the beam penetration depth (~5 µm in sapphire at 35 keV).

Efforts have also been made at obtaining lattice parameters and elastic strains from ECPs. High order lines are more sensitive to changes in lattice parameter (strain) but are more difficult to detect in ECP due to the reduced contrast and smaller line width. No general method has been demonstrated but it is often possible to find regions of the pattern that contain lines that are quite sensitive to lattice parameter (strain). The best accuracy that has been quoted is ±3 parts in $10^4$ obtained by Walker and Booker (1982) for lattice parameter measurements in Si and GaP using lines such as {10,10,0} which are not routinely visible in ECP. Other workers have obtained strain sensitivities of ±2 parts in $10^3$ (Kozubowski, Keller and Gerberich 1991) but even here the {660} lines used would be readily removed by any plastic deformation in the crystal.

The ECP technique is available on many SEMs and is likely to remain in use on an occasional basis in many laboratories, mainly for orientation measurement. However the main limitation of the technique is the spatial resolution which even at its best of ~2 µm, is an order of magnitude worse than that offered by the EBSD technique.

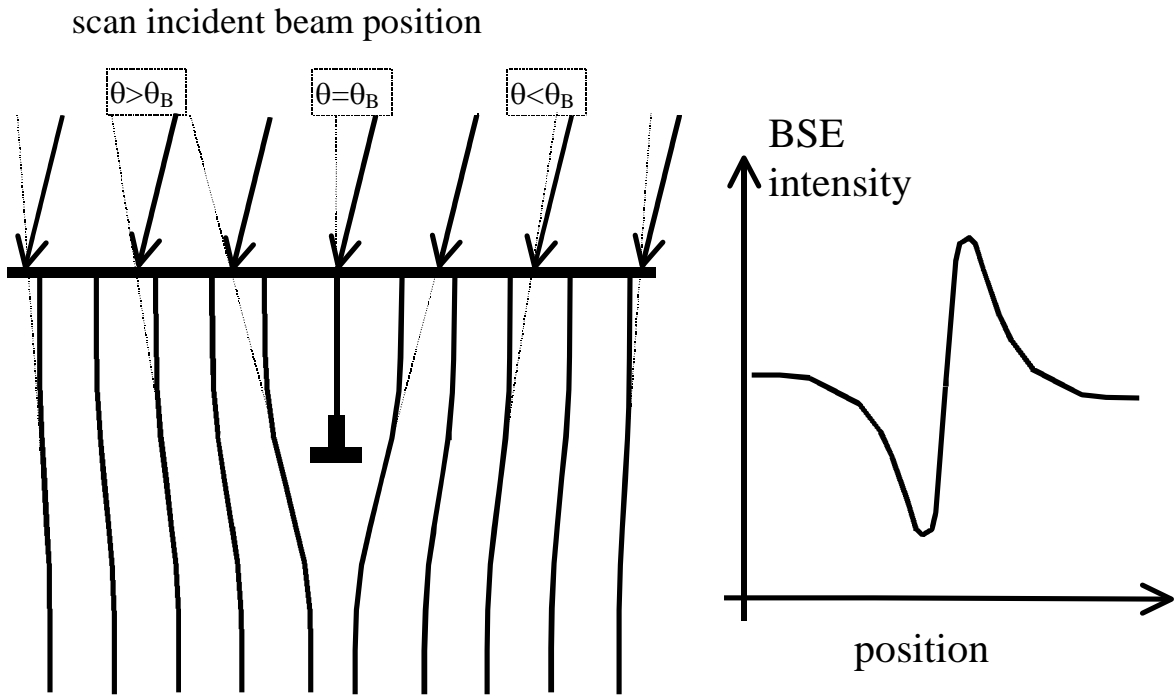

*Fig. 9 Schematic diagram of the ECCI technique showing how the local lattice plane tilting near a dislocation causes a modulation of the BSE intensity allowing the defect to be imaged.*

## III. ELECTRON CHANNELLING CONTRAST IMAGING OF CRYSTAL DEFECTS

### III A  Imaging Dislocation Lines

*(a) prediction of electron-optical conditions required*

The basic idea of using channelling contrast to image dislocations was put forward in the very early paper by Booker *et al* (1967), who stated that 'it was *only* necessary to orient the crystal at the Bragg position and the local bending of the crystallographic planes where the dislocations emerge at the surface should provide the necessary contrast'. The idea is illustrated in fig. (9) which shows an incident beam being scanned across a crystal that contains a dislocation close to the specimen surface. The incident beam is at the Bragg condition for the good crystal away from the dislocation, but as the beam is brought close to the dislocation the lattice plane tilting on the left of the dislocation gives a local positive deviation from the Bragg angle and so we expect (from fig. 4) that the BSE intensity will be locally reduced. Conversely to the right of the dislocation the lattice plane tilting leads to negative deviations from the Bragg condition and hence a local increase in BSE intensity. The sense or directionality of the contrast is controlled by the Burgers vector of the dislocation and as initially noted by Booker *et al* (1967) can be used in defect characterisation. The first assessment of the likely electron-optical conditions required to image dislocations was given by Booker (1970) who realised that the local tilting was of the order of $10^{-3}$ rad at 30 nm from the dislocation and fell to $10^{-4}$ rad at 300 nm, so that it would necessary to keep both the beam divergence and the probe diameter small, while the limited contrast available would be likely to lead to restrictions based on the signal to noise ratio.

More detailed calculation of dislocation image profiles followed with Clarke and Howie (1971) using a single scattering model to get the form of the image and an empirically determined constant term to account for the multiply scattered background. Spencer, Humphreys and Hirsch (1972) used the forward-backward approximation in an attempt to calculate the image contrast directly. The results of these calculations were in broad agreement and example image profiles of screw dislocations calculated using a simplified form of the Spencer *et al* (1972) model developed by Wilkinson *et al* (1993a) are given in fig. (10). At the Bragg condition in the symmetric Laue case the images of dislocations close to the specimen surface have the bright-dark form suggested in fig. (9) when **g.b**=1, while extra oscillations are introduced near the dislocation when **g.b**=2, or indeed takes higher values. The calculations for fig. (10) assumed a dislocation parallel to and 10 nm below the surface of a bulk Si specimen imaged with a {220} reflection. Since the extinction distance changes with beam energy the beating pattern of the Bloch wave field moves relative to the fixed dislocation, and this causes the reversal of the sense of the dark-bright contrast, the dislocation depth being $0.13\xi_g$, $0.22\xi_g$ and $0.52\xi_g$ at energies of 100, 30 and 5 keV respectively. Such behaviour was shown in the simulations made by Spencer *et al* (1972).

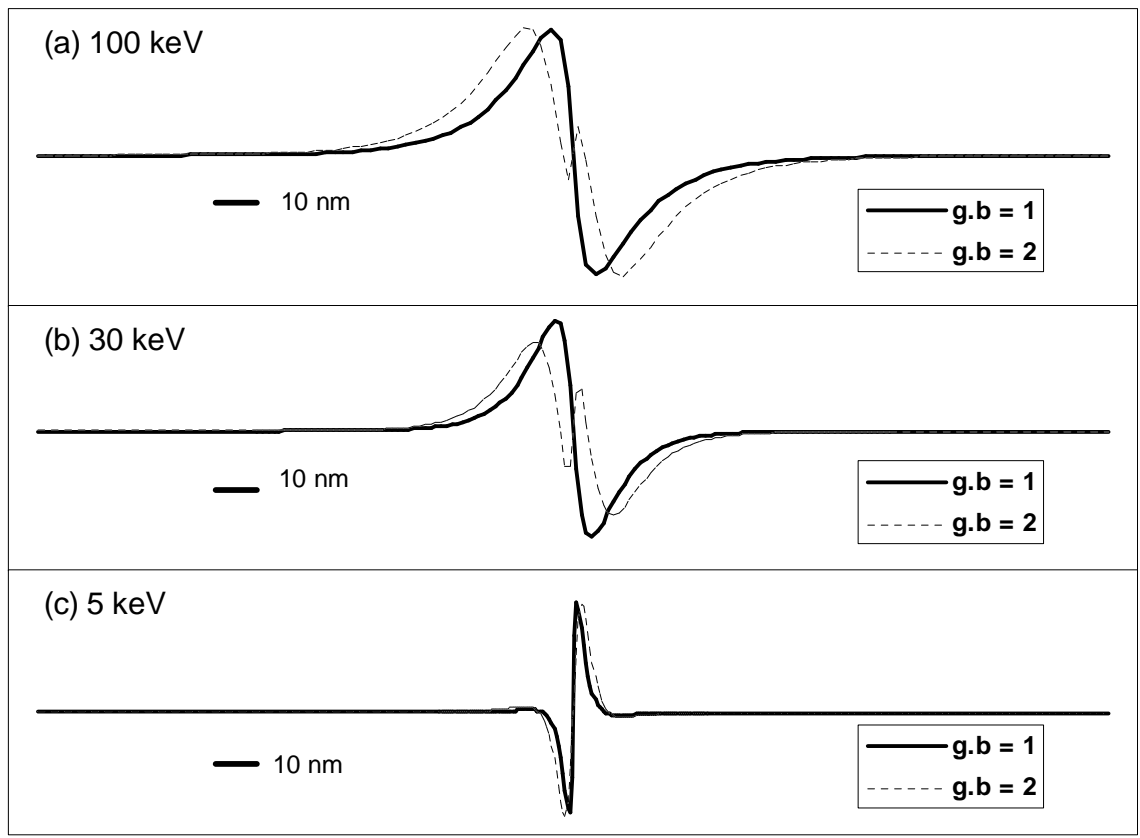

*Fig. 10  Simulated ECCI intensity profiles for a screw dislocation parallel to and 10 nm below the surface of a Si specimen, imaged at normal incidence using a {220} reflection. Note the decrease in image width at lower beam energies. Image dislocation lines included.*

It is important to note that the image widths in fig. (10) are narrow and decrease as the extinction distance ($\xi_g$) decreases with beam energy from 100 keV down to 5 keV. Indeed, for a dislocation $0.2\xi_g$ below the surface, Spencer *et al* (1972) suggested that the probe diameter ($d_0$) must be kept below a limit of ¼ $\xi_g$ **g.b** so as to ensure that the majority of the available contrast is retained experimentally. At 30 keV this corresponds to a beam diameter of less than ~10 nm for low order reflections in the majority of crystals. The need for a finer beam at lower beam energies, is of course opposed by the increase in beam diameter arising from aberrations in the magnetic lenses used in SEMs.

The images are also highly sensitive to deviation from the Bragg condition (if this were not so then of course the dislocations could not be imaged), with the contrast in the image falling off quite drastically for rotations as small as a few mrad away from the Bragg angle. A consequence of this, with important practical implications, is that the finite beam divergence used experimentally will average the image over a range of incident beam conditions and so reduce the image contrast. This effect is illustrated in fig. (11) again using results calculated with the model of Wilkinson *et al* (1993a) over a range of beam divergences α, where the total angle subtended by the beam is 2α. Spencer *et al* (1972) suggested that in practice it would be necessary to ensure that the total beam divergence was less than the angular width of the channelling line for the reflection selected. In the two beam theory the channelling line width $W_g$ is given simply by $W_g = 2/|\mathbf{g}|\xi_g$ ,thus $\alpha < 1/|\mathbf{g}|\xi_g$, which for a typical reflection (Si 220) at 30 keV corresponds to an upper limit on α of 4 mrad. Since $\xi_g$ increases with the velocity of the electron, the limiting value of α becomes smaller at higher energies, and this can be seen in the simulations given in fig. (11), where α is seen to affect the image contrast more at 100 keV than 5 keV.

These limits on the diameter and divergence of the beam restrict the current that is available at the specimen to form the image. Indeed combining Spencer's expressions for $d_0$ and α with the relation ($B=4I_0/\pi^2\alpha^2 d_0^2$) between the source brightness (B) and the current $I_0$ at the specimen, a condition for the brightness can be established where $B > 6.4\, I_0 / b^2$, with b the component of **b** along **g**. The problem that this makes apparent is that the current at the specimen ($I_0$) must be kept large enough to make the defect visible over the background noise, which can only be achieved if the brightness is greater than some critical value. An often used criterion for visibility of a feature in an image is that $C\sqrt{n} \geq 5$, where C is the contrast of the feature and n is the average number of electrons detected at each point in the image. The number of electrons detected clearly depends on the current present in the beam, and so this visibility criterion sets the minimum current which is required at the specimen. In fact $n = I_0\, \eta\, \varepsilon\, t_f / m\, e$, where η is the backscatter coefficient, ε the detection efficiency, $t_f$ the frame time, m the number of picture points and e the electronic charge. Combining this expression with the previous two inequalities produces an expression for the minimum source brightness $B_c$ required in order to obtain well resolved clear images of near surface dislocations,

$$B_c = 160\, m\, e\, /\, b^2\, C^2\, \eta\, \varepsilon\, t_f \qquad\qquad (1).$$

Clearly the image contrast is an important parameter in determining the minimum brightness required. Spencer *et al* (1972) noted that the contrast in the simulated dislocation images was similar to that calculated for the channelling line used, and thus is expected to be small. Furthermore Spencer *et al* argued that the image contrast would decrease with increasing beam energy. Experimentally the contrast of ECPs has been shown to decrease with beam energy; Yamamoto measuring a contrast of ~5% at 20 keV falling to ~2.5% at 50 keV for a {220} band in Si. Although the channelling contrast is greater at lower beam energies this must be considered against the decreasing brightness of the electron source and the decreasing performance of BSE detectors, which tends to give a marked deterioration in terms of signal to noise ratio for beam energies below about 10 keV.

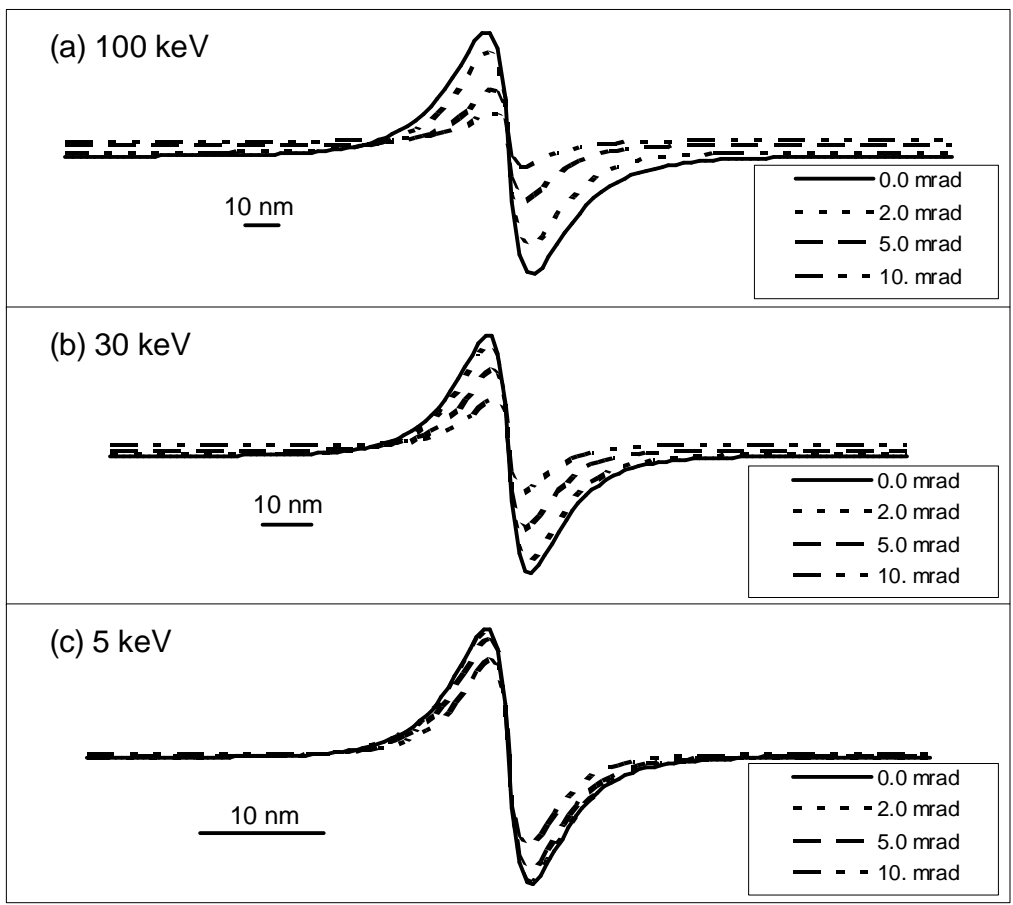

*Fig. 11  Effect of the beam divergence α on simulated ECCI intensity profiles for a screw dislocation parallel to and $0.2\xi_g$ below the surface of a Si specimen, imaged at normal incidence using a {220} reflection. Note that beam divergence has a larger effect at higher beam energies. Image dislocation lines included.*

Inserting the relevant numerical values into equation (1) we find that the minimum brightness needed at 30 keV is ~$1 \times 10^{12}$ Am$^{-2}$srad$^{-1}$. This brightness cannot be achieved with thermionic emission sources, such as the commonly used W hairpin, or LaB$_6$ single crystals. Instead an SEM with a field emission gun (FEG) is required, and even such a system is close to the detection limit.

*(b) experimental developments*

Initial attempts made in the early 1970s to image dislocations using ECCI were limited to using the W thermionic emitters then available. Clarke (1971) made some observations in a TEM fitted with a scanning attachment so as to make use of the improved brightness of the filament when operated at 80 keV. By working with a thin specimen the effects of multiple scattering could be reduced considerably so decreasing the background signal level and hence increasing the contrast. The specimen was oriented to show dislocations with clear contrast in a scanning transmission image and an image then obtained using BSEs collected by a detector subtending ~1.2 srad at the specimen. Images of dislocations were obtained from a heavily deformed Al specimen and images that Clarke identified as stacking faults were obtained in Cu-8% Al alloy. Images were also obtained by Stern *et al* (1972) using thin foils of molybdenite in which arrays of dislocations lying in the basal plane could be seen again with transmission and BSE images obtained from thin foils. Further and more detailed observation were made in thin foils by Booker *et al* (1973) who showed that the contrast of dislocation lines depended on their depth in the foil and on deviation from the Bragg condition. However the BSE images were of poor quality and could only be obtained from thin parts of the specimen, so that the technique had no advantage compared to conventional TEM.

In the late 1970s a great improvement in the image quality was made. Pitaval *et al* (1975) demonstrated that better quality ECCI images of dislocations in thin molybdenite foils and stacking faults in thin stainless steel specimens could indeed be produced by using a high brightness FEG which they had fitted to an SEM. The FEG was operated at a temperature 1200 K and at 50 keV an emission current of 100 µA lead to a brightness of $3 \times 10^{12}$ Am$^{-2}$srad$^{-1}$. Even so, in thicker parts of the specimens the channelling contrast was reduced and the defect images could not be produced. In a subsequent publication (Pitaval *et al* 1977) they also described the construction of a retarding field energy filtered BSE detection system with which they showed that energy filtering could enhance contrast of ECPs and remove contrast asymmetries and reversals that occur in unfiltered ECPs from tilted crystals. Most importantly, they showed for the first time that dislocations could be imaged in thick molybdenite crystals by using the FEG operating at 45 keV, in conjunction with the energy filter rejecting BSEs that had lost more than 1 keV through inelastic scattering. In later publications (Morin *et al* 1979a, Morin *et al* 1979b , Fontaine, Morin and Pitaval 1983) the image quality was further improved by setting the energy window of the detector to reject electrons losing 500 eV or more so as to optimise the signal to noise ratio. Despite the impressive images of dislocations, stacking faults, and microtwins that were obtained the technique was not adopted by other groups. The reasons for this could include the need for a specialised FEG SEM instrument with an energy filter, and the great interest in development of high resolution transmission electron microscopy.

The next development of the technique did not occur until a decade later when Czernuszka *et al* (1990) obtained images of dislocations in bulk specimens using a FEG STEM instrument operated in scanning mode without the need for an energy filtered detector. This method has since been used successfully in a conventional FEG SEM (Wilkinson and Hirsch 1995) with very little modification to an instrument which is widely available commercially. Joy (1990) has also produced images of dislocation arrays in molybdenite using a FEG SEM in which the specimen is immersed in the magnetic field between the pole pieces of the objective lens. This lens configuration gives smaller aberrations but restricts the size and manipulation of the specimen compared to the more conventional snorkel lens design that is more widely used in SEMs.

Central to the success of the ECCI technique, as employed at Oxford without an energy filtering detector, has been the adoption of a highly tilted specimen geometry with a detector positioned so as to collect BSEs scattered into low takeoff angles (shown in fig. 12). This acts to improve the signal to noise ratio in three ways. Firstly, the yield of BSEs increases as the specimen is tilted away from normal incidence as shown for example by Drescher, Reimer and Seidel (1970), or Darlington and Cosslett (1972) so that there are more electrons providing the signal (i.e. $\eta$ is increased in equation 1). Secondly, there exists a pronounced peak near the forward scattered specular direction in the angular distribution of the BSEs as was shown by Seidel (1972) and Wells (1975) so that the electrons can be detected quite efficiently. A detector placed to intersect this forward scattered peak therefore collects a large proportion of the BSEs that are emitted (i. e. $\epsilon$ is increased in equation 1). Thirdly, the energy distribution of BSEs emitted from a tilted specimen has been shown by Matsukawa, Shimizu and Hashimoto (1974) to have a larger proportion of electrons at the high energy (low loss) end of the spectrum compared to the case at normal incidence. Indeed the energy peak becomes progressively larger and moves further toward the low loss end as the specimen is tilted, the effect being quite marked for low to medium atomic numbers. The tilting thus acts to reduce the amount of multiple scattering and results in the channelling contrast being enhanced as was shown by Yamamoto (1977) who measured the absorbed current signal to deduce the channelling contrast in the total BSE signal. In practice the level of channelling contrast depends on the position and size of the detector and its energy response. Indeed Matsukawa *et al* (1974) showed that for an Al specimen tilted to $\theta=60°$ the energy distribution measured at low take off angle ($\phi=10°$) shows a very large peak at the low loss end which is about 5 times higher than the peak measured with a detector mounted approximately normal to the specimen surface.

Fig. (12 a) shows the specimen and detector configuration currently used for ECCI investigations at Oxford. The specimen is tilted from 40° to 70° away from normal incidence, as dictated by the selected channelling plane, and the signal from a retractable BSE detector held at low take off angle is used to form the ECP or image. This detector consists simply of a ~12 mm diameter YAG scintillator, mounted on a light guide which couples it to a photomultiplier tube (PMT) outside the microscope. The YAG scintillator is coated with Al, firstly to filter out contributions from lower energy electrons, secondly to prevent charging, and finally to reflect emitted light into the light guide and towards the PMT. This assembly can be translated towards or away from the specimen on a horizontal axis just below the specimen height. Translation sweeps the detector through a range of takeoff angles and from the rise and fall of the detected signal level the position (B in fig 12a) corresponding to the forward scattered peak in the BSE distribution can be found.

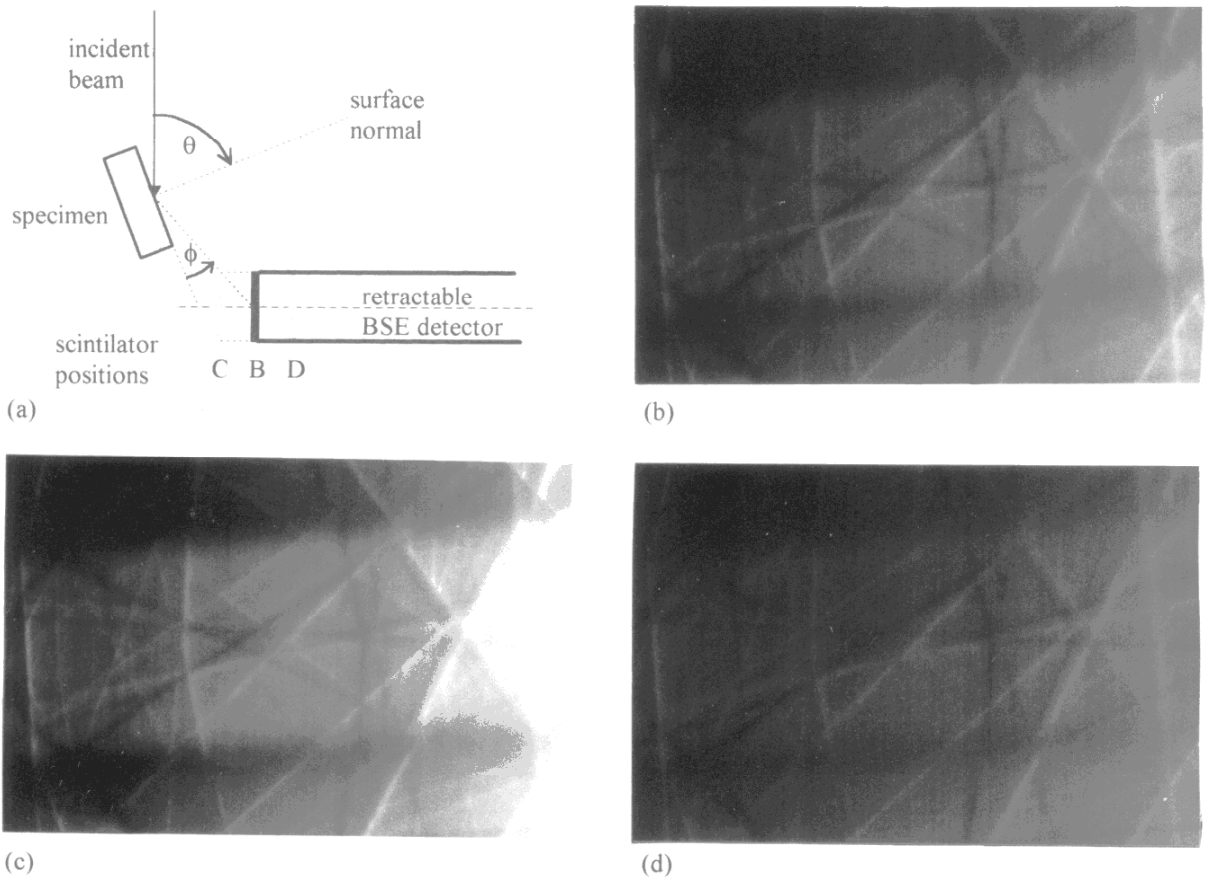

*Fig. 12*     *(a) shows the geometry used for ECCI observations with the specimen highly tilted (tilt angle defined as θ), and a moveable BSE detector positioned at low takeoff angle (takeoff angle defined by ϕ). The detector at position B corresponds to the maximum signal level, and the ECPs shown in (b), (c) and (d) were obtained with the detector in positions B, C, and D respectively.*

Although the signal level is at maximum at B the signal to noise ratio for channelling contrast is not. Fig. (12) (b), (c), and (d) show ECP at 30 keV from a Si specimen tilted 70° from normal incidence, with the ECCI detector at positions B, C and D respectively, as indicated in fig. (12a). The signal to noise ratio is best with the detector positioned at a lower takeoff angle than that giving the peak signal. This is in accord with experiments showing more low loss electrons scattered to low takeoff angles and with the calculations of Dudarev *et al* (1995).

Using this tilted specimen, low takeoff angle detector configuration dislocations have been imaged in bulk specimens using ECCI. Several microscopes were evaluated and fig. (13) shows dislocation images obtained from some of them. For comparison the images were all taken from a single crystal Si specimen deformed in compression along [123] at 450°C and then sectioned on the primary (111) glide plane, which contains dislocations with line directions along the three <110> directions contained in the glide plane. The first two images were obtained by Czernuszka *et al* (1991) using a VG HB501 FEG STEM operated in scanning mode with the specimen at a long working distance outside the objective lens. The difference in image quality obtained at 100 keV (fig. 13a) and 30 keV (fig. 13b) is striking. The signal to noise ratio is clearly better at the lower beam energy despite the lower beam current available. Estimations of the signal to noise ratios from dynamical diffraction theory calculations show that the clearer image obtained at 30 keV is a result of the increase in channelling contrast with reduction of the beam energy (Wilkinson *et al* 1993b). Image widths are noticeably larger at the higher beam energy which is in agreement with theory (see fig. 10). Despite the poorer channelling contrast obtained at 100 keV, an advantage is that the longer extinction distance leads to dislocations being imaged to greater depths. Indeed Czernuszka *et al* (1991) imaged dislocations in a Si specimen polished 6° off the (111) glide plane so as to estimate that dislocations could be imaged to a depth of ~200 nm at 100 keV and about half that depth at 30 keV.

Images have also been obtained with a JEOL JSM 840F (fig. 13c), demonstrating that the technique can used on the more widely available commercial FEG instruments (Wilkinson and Hirsch 1995). The image width in fig (13c) is ~35 nm at its narrowest which agrees well with theory (see fig. 10), while the broader parts of the dislocation images which generally also show lower contrast are thought to occur for dislocations deeper into the crystal. The dislocation contrast is seen to vary from segment to segment in fig. (13c) sometimes being bright-dark and sometimes predominantly bright or predominantly dark. The early calculations by Spencer *et al* (1972) showed that bright-dark images of near surface dislocations became predominantly bright as the dislocation depth was increased due to exhaustion of the more highly scattered Bloch wave so that at larger depth electrons can be scattered into but not from this Bloch wave. The observation of predominantly dark dislocation images was explained by Wilkinson and Hirsch (1995) for the case of highly tilted specimens where in addition to the depth variation other derivatives of the displacement field make significant contributions to the contrast.

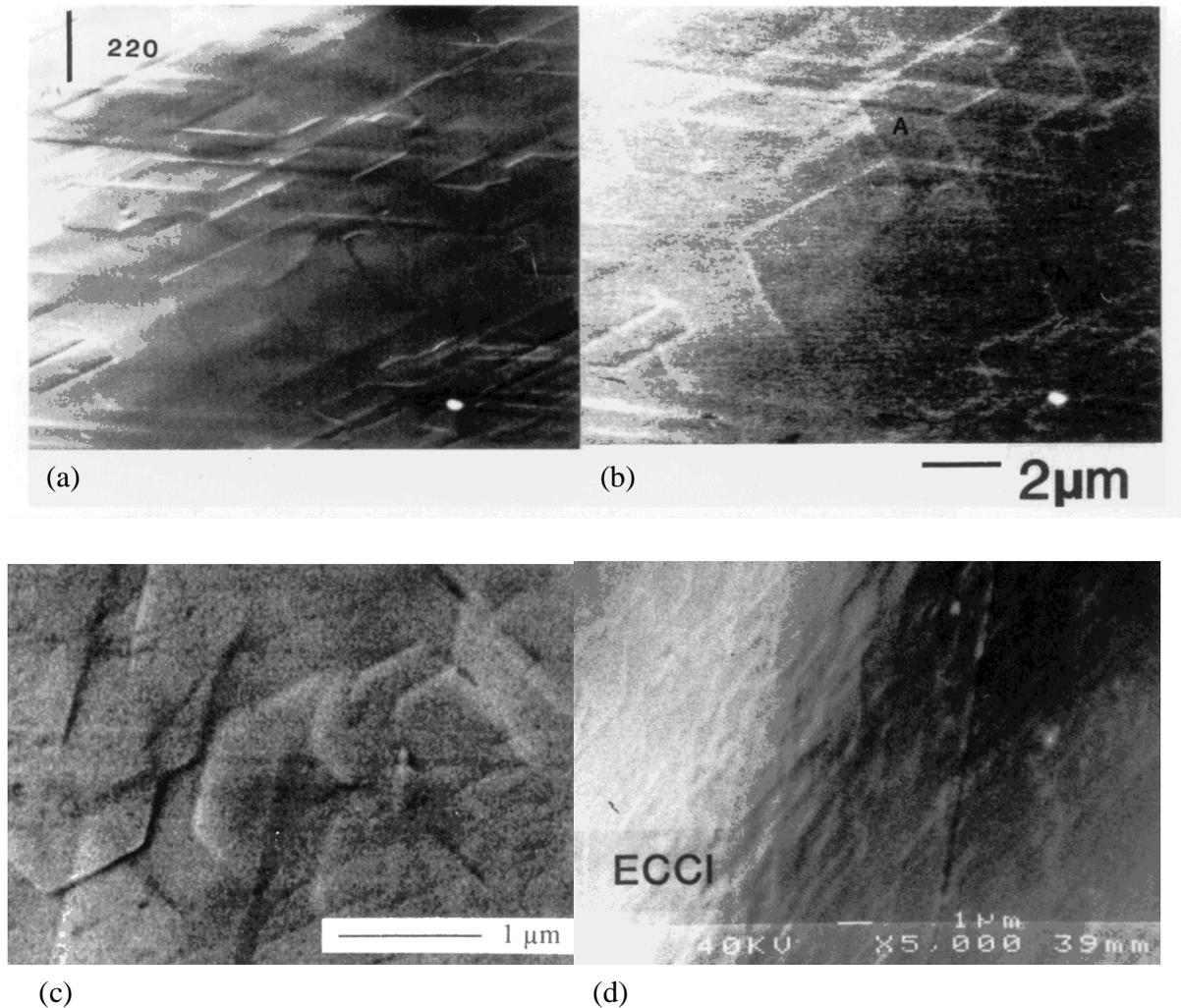

*Fig. 13    ECCI showing dislocations in a bulk Si specimen. (a) and (b) were obtained by Czernuszka et al (1991) in a VG 501 FEG STEM at beam energies of 100 keV and 30 keV respectively. (c) was obtained in a JEOL 840F FEG SEM at 35 keV (Wilkinson and Hirsch 1995), and (d) in a JEOL 6300 with a LaB$_6$ filament operated at 40 keV (Wilkinson et al 1993a).*

The condition given in equation (1) indicates that a LaB$_6$ filament is not sufficiently bright to produce well resolved images of individual near surface dislocations. However, fig. (13d) obtained with a JEOL JSM 6300 at 40 keV with a LaB$_6$ filament (Wilkinson *et al* 1993a), shows that, although not well resolved, dislocations can be just detected. The images are of low contrast and diffuse, with the image widths of ~250 nm being much larger than obtained with a FEG SEM.

*(c) characterisation of dislocations*

In addition to imaging dislocation distributions in bulk specimens the ECCI technique allows the character of the dislocations to be assessed through the nature of the contrast observed under different diffraction conditions. The simplest case is of course for screw dislocations where the **g.b**=0 invisibility criterion, established using diffraction contrast in the TEM, can be used directly since there are no displacements perpendicular to the dislocation line. Czernuszka *et al* (1990) showed this extinguishing of ECCI image

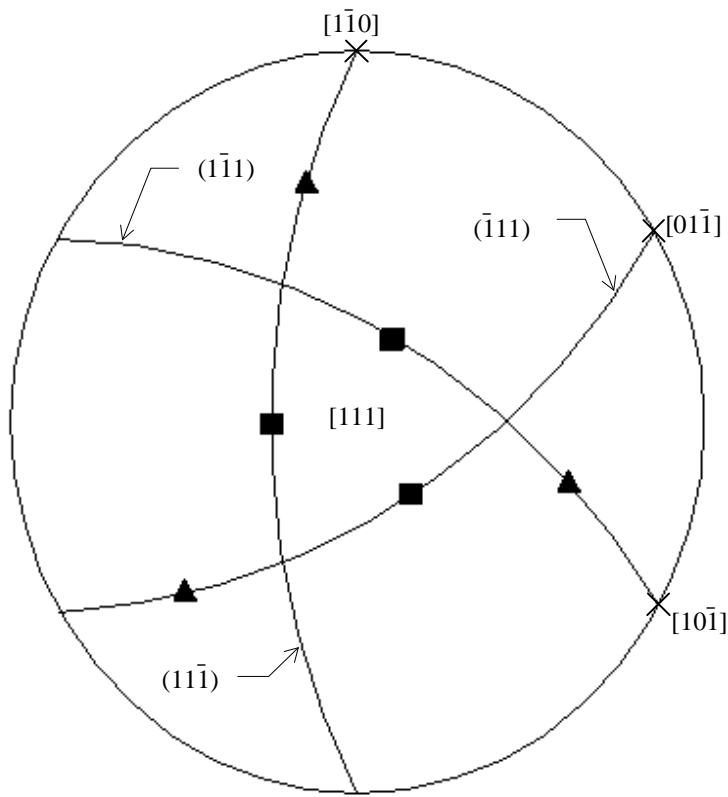

*Fig. (14a)*

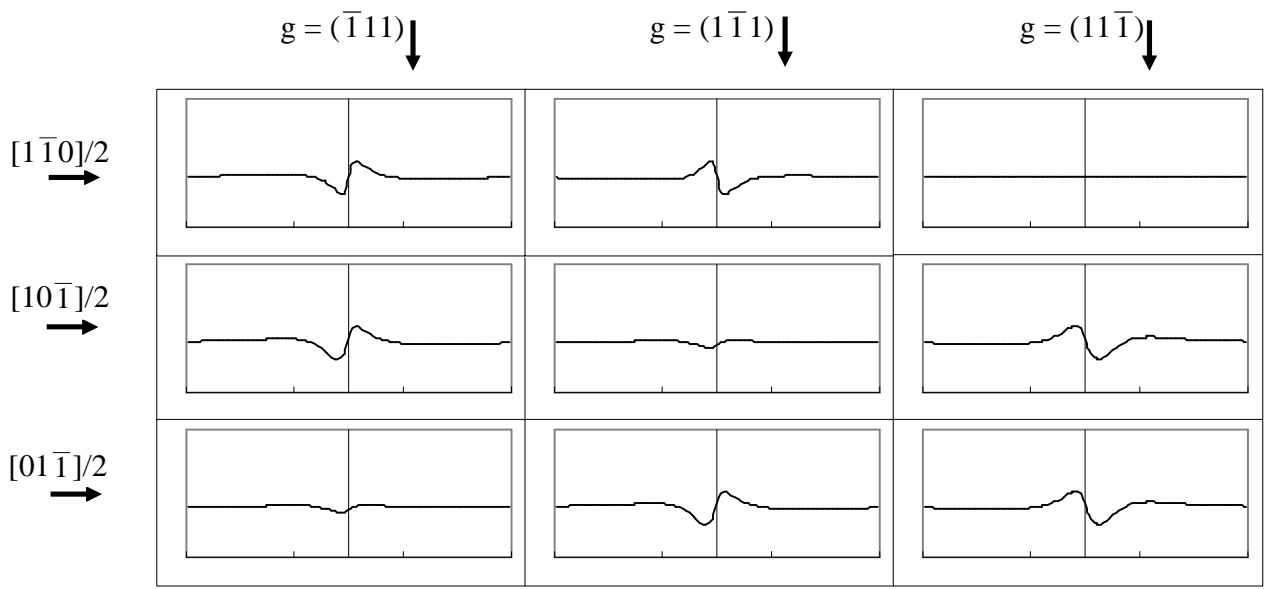

Fig. (14b)

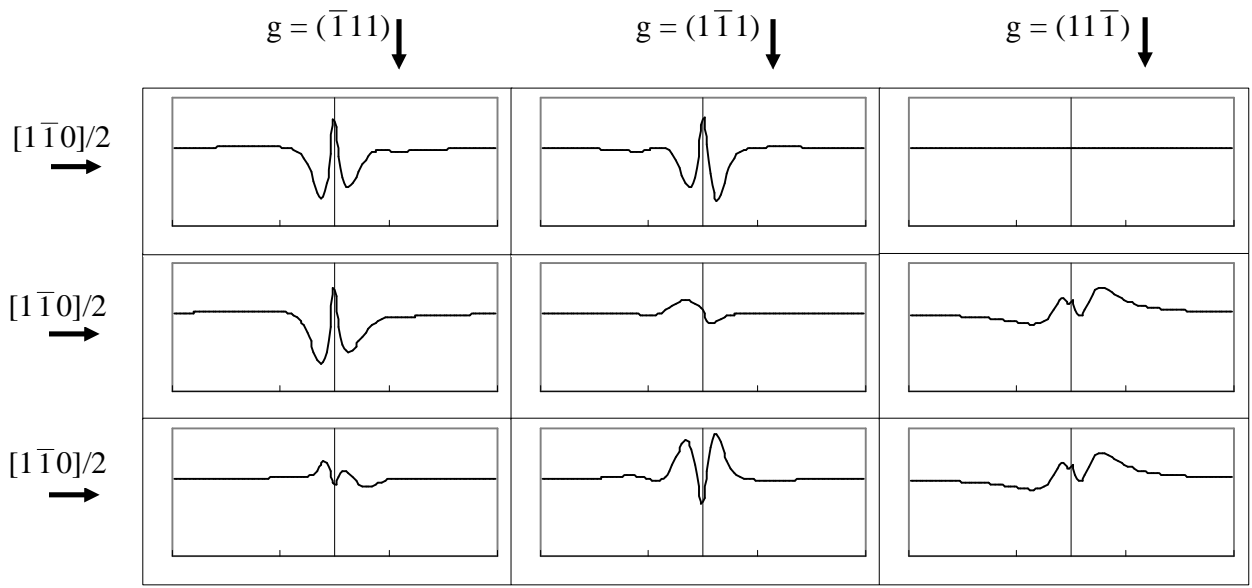

Fig. (14c)

*Fig. 14*   *(a) A stereogram centred on the specimen surface normal which is the (111) glide plane on which glissile Burgers vectors for dislocations along [1$\bar{1}$0] are indicated as crosses, together with the {111} diffraction planes, and incident beam directions used in defect characterisation. (b) Simulations of ECCI intensity profiles for dislocations with the indicated Burgers vectors imaged under the three indicated {111} reflections with the incident beam inclined 19.5° from normal incidence, as indicated by the filled squares in (a). (c) Similar simulations but with larger inclination of the incident beam as indicated by the filled triangle in (a). (from Wilkinson and Hirsch 1995)*

contrast that occurs when **g.b**=0 using a deformed Ni$_3$Ga single crystal known from TEM observations to contain screw dislocations (Sun 1990). With edge and mixed dislocations the displacements are not confined to a single direction and so fulfilment of the simple **g.b**=0 criterion does not extinguish all the image contrast. However, in TEM such analysis can still be used since the contrast is generally weaker when **g.b**=0 (see Hirsch *et al* 1978 pages 181-182). The calculations of Wilkinson and Hirsch (1995) have shown that it may be more difficult to characterise dislocations using ECCI from highly tilted specimens. They considered a Si specimen sectioned on a (111) glide plane containing three possible glissile dislocations and simulated images of them obtained with three {111} reflections (see fig. 14). In TEM these reflections would be accessed using a small sample tilt of ~20°, and ECCI images under these conditions (fig. 14b) would allow the different dislocations to be distinguished due to the weak or zero contrast obtained with one of the reflections. However, in practice ECCI is undertaken at higher specimen tilts and under these conditions the difference between strong and weak contrast appears to be less marked (fig. 14c). However, Morin *et al* (1979a) have demonstrated this weakening of dislocation contrast when **g.b**=0 in experiments carried out with their energy filtering detector using 220 reflections in Si, as shown in fig. (15).

A further geometrical aspect that can influence the observed image contrast and hence is important when assessing contrast levels for a Burgers vector determination is the dislocation line direction. The angle between the dislocation line and the incident beam direction can vary significantly during ECCI observations of a highly tilted specimen using different reflections. The importance of this angle was first noted by Wilkinson *et al* (1993a) during observation on groups of long straight misfit dislocation in strained epilayer structures which will be discussed in more detail in section III B. Figure (16) taken from this work shows how the dislocation contrast varies as the crystal is rotated through a large angle while maintaining the same operating reflection, namely **g** = 022. The contrast is strongest when the line direction is close to perpendicular to the incident beam and steadily decreases as the rotation brings the line direction closer to the beam direction. The effect arises since the contrast results from the derivatives of the displacement field along the incident beam direction, and since the displacements are constant along the dislocation's length the derivatives are reduced as the line direction is brought closer to the beam direction. This effect is well known in TEM (Hirsch, Howie and Whelan 1960) and is of importance in ECCI when specimens are highly tilted.

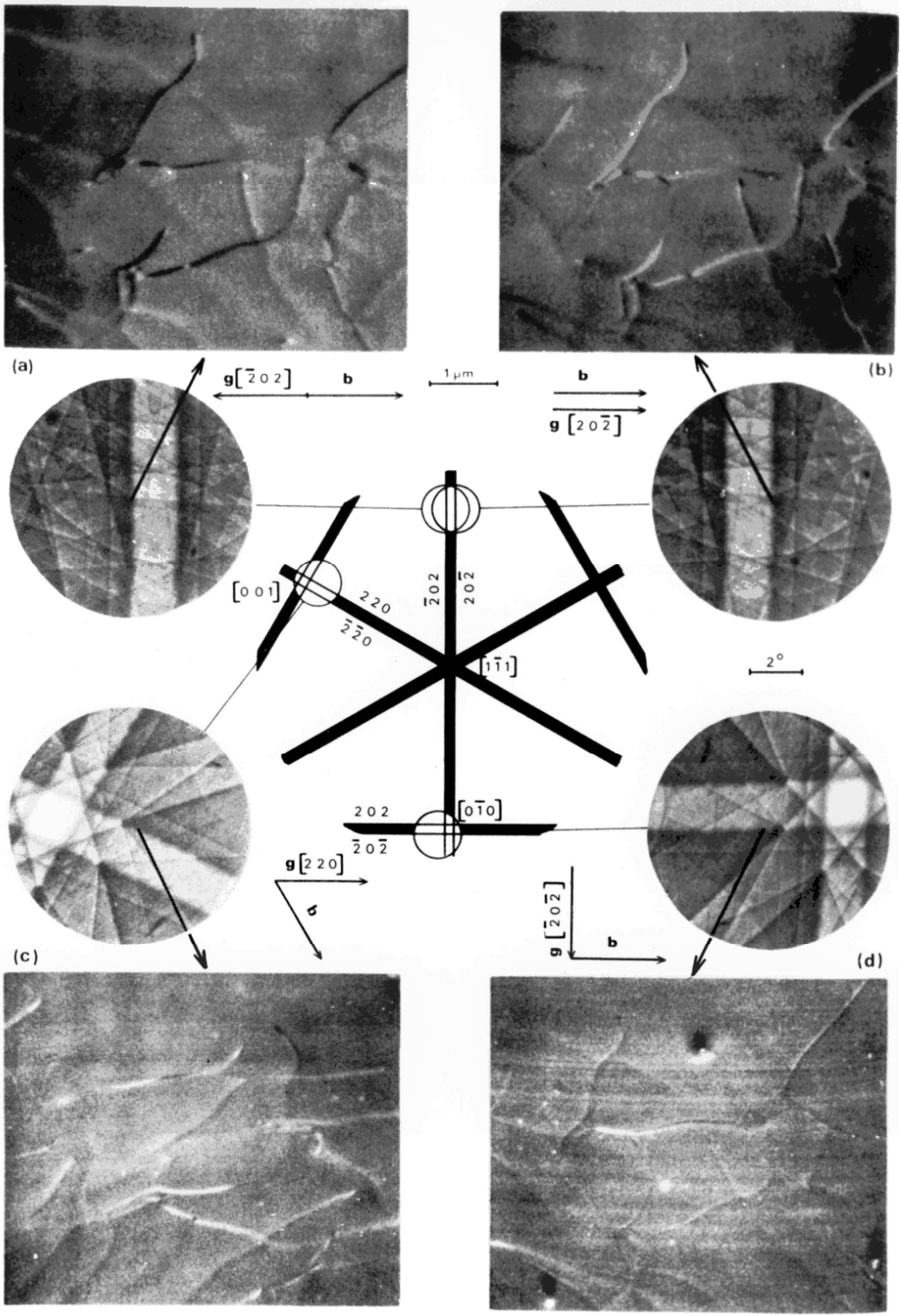

*Fig. 15*     *Images of dislocations in Si obtained under various diffraction conditions by Morin et al (1979) at 45 keV, with an energy filtered detector. Dislocation images in (d) obtained with **g.b**=0 show weaker contrast than in the other images where **g.b**≠0. (from Morin et al 1979)*

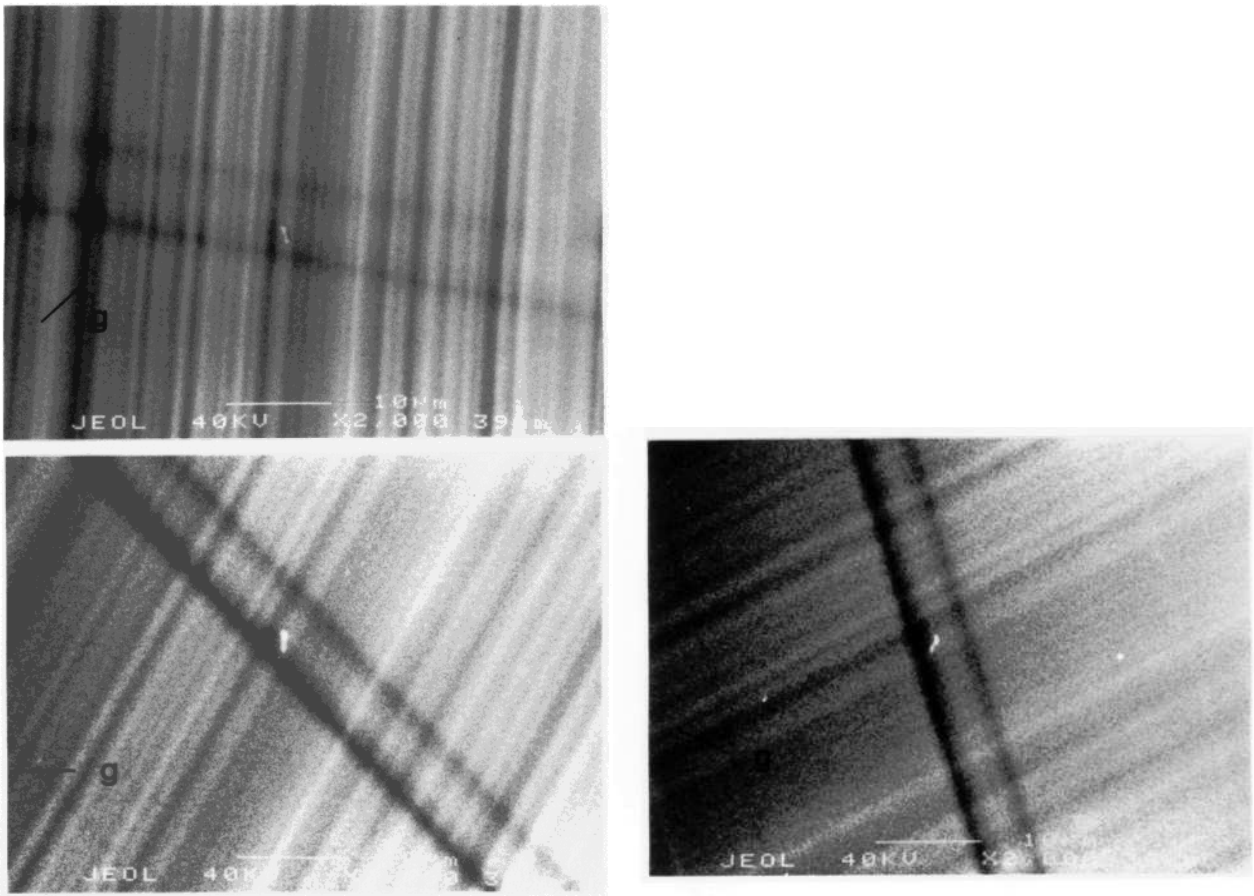

*Fig. 16    ECCIs from the same region of a 0.72 μm thick $Si_{0.95}Ge_{0.05}$ epilayer on a Si substrate all taken with g = 022, but with different incident beam directions. The defect contrast is steadily decreased as the specimen is rotated to bring the line direction closer to the incident beam direction. The tilt axis is approximately vertical in each image. (from Wilkinson et al 1993)*

### III B  Long Range Strain Fields

Although dislocations cannot be well resolved using a SEM with a $LaB_6$ thermionic emitter, an effect that is of importance is shown in fig. (13d). In addition to the contrast modulations seen at the 250 nm scale which are due to dislocations, a broader contrast variation across the whole image from bright on the left to dark on the right can also be seen. This contrast moves as the specimen is translated and so is not a small region of a Coates's Kikuchi-like reflection pattern. The effect arises from the overall, bending in the bulk crystal due to the distribution of dislocations that it contains. Such long range strain fields are of importance in many aspects of materials science, and ECCI using a SEM with a thermionic emitter is a suitable means of studying them.

Figure (17) compares the imaging of a single near surface dislocation with imaging of a group of dislocations lying deeper within the crystal. The information depth for channelling contrast is at most 100 nm as beyond this the initial Bloch wave states are exhausted. If a dislocation lies within this depth the image width is small and a FEG SEM is required for imaging, as was discussed in section III A. For the deeper lying defects although they are beyond the information depth, each dislocation causes some distortion of lattice planes close to the surface. These long range strains from each line overlap and so ECCI can be used to probe their combined strain fields. In the case shown in fig. (15b) the long range strain from the distribution of dislocations can be approximated quite well by replacing the group with a single dislocation with the same net Burgers vector. Since the strain is sampled at a large distance from the defects, its variation occurs over a longer distance at the specimen surface. This drastically reduces the restrictions that were placed on the beam diameter for imaging individual dislocations, so that the brightness requirements are less severe. In fact, in these cases thermionic emitters become the preferred electron sources since they allow greater beam currents, and hence better signal to noise ratios to be employed than are possible with FEGs. Some examples of observations that can be made with the ECCI technique at lower spatial resolution are given below.

 *(a) Observations of Strained Epilayers*

During the growth of strained epitaxial layers the stored elastic energy increases until it becomes favourable for some of the strain to be accommodated plastically by the generation of misfit dislocations at the interface. The dislocations are often found to be in bunches of 5 to 10 lines separated by distances that can be as small as a 10 nm. Since the epilayer thickness can be orders of magnitude greater than the width of the bunch of dislocations, the situation is very much like that illustrated in fig. (17b). Example images of misfit dislocations in $Si_{1-x}Ge_x$ grown on Si have already been given in fig. (16), from which we can see that the image width is of the order of 1 µm, for this 0.72 µm thick epilayer, which is markedly wider than the near surface dislocation images produced with the same instrument. Image simulations (Wilkinson *et al* 1993a) showed for these deep lying defects the image width increases linearly with depth, and experimental intensity profiles given by Wilkinson *et al* (1994) confirm this predicted trend.

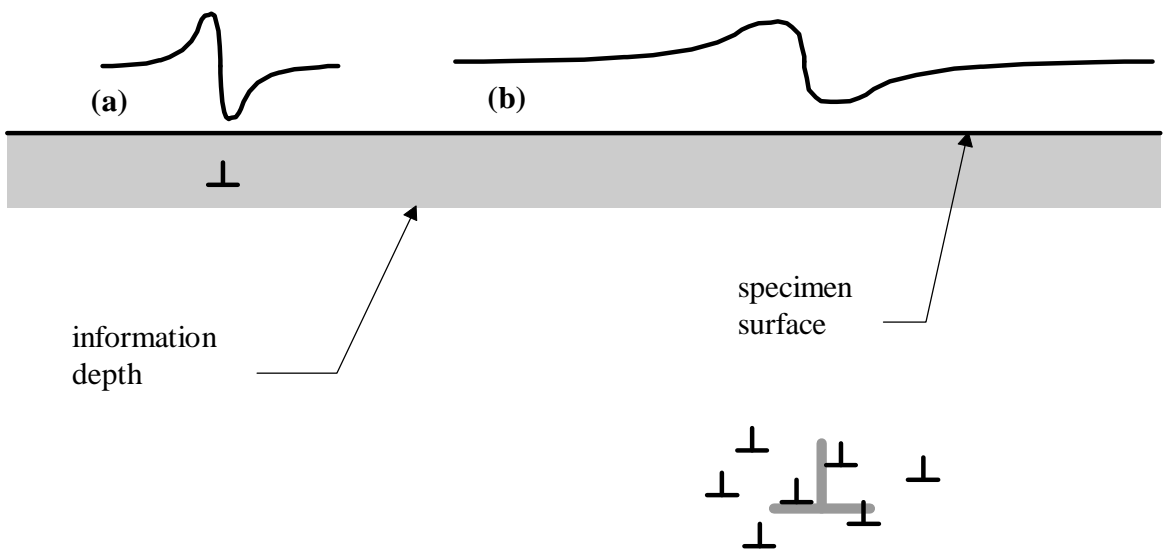

*Fig. 17      Geometry leading to (a) narrow image widths for individual dislocations within the information depth and (b) wider images from the overlap of long range strain fields from a group of dislocations deeper with the crystal.*

The ECCI technique very quickly reveals the distribution of defects in such epilayer structures, and large areas can be assessed directly in the as grown material. This fact lead Wilkinson (1996a) to try relating the dislocation distributions seen in $In_{0.2}Ga_{0.8}As$ epilayers grown to various thicknesses on GaAs to the macroscopic strain relaxation. For the thinnest epilayer (25 nm thick) individual dislocations and clearly demarked bunches of dislocations were observed, but for thicker epilayers the increasing defect content resulted in complicated fluctuations in the image contrast. Fourier analysis of intensity traces taken from ECCI images were used to assess whether the strain fields from the defects overlapped significantly over the entire interface. This was found to happen first at an epilayer thickness of 100 nm which corresponded well with the onset of macroscopic relaxation as detected by x-ray diffraction measurements.

Keller and Phelps (1996) have also managed to image misfit dislocations in $In_{0.25}Ga_{0.75}As$ on GaAs, producing images similar to those obtained by Wilkinson (1996a). Again the specimen is highly tilted (70°) and a scintillator detector placed at low takeoff angles, though their detector has a markedly smaller capture angle than in the system used by Wilkinson (1996a). The smallest image width they reported was ~90 nm for a 100 nm thick epilayer.

Characterisation of the dislocation bunches is complicated since although the individual lines have <110>/2 type Burgers vectors, the overall Burgers vector need not be. Attempts were made to verify that the net Burgers vector was such that the defects acted to relieve the compressive in-plane stress within the epilayers, from the nature of their observed image contrast. This led to the realisation that surface stress relaxation plays an important role in determining the contrast from these deep lying dislocation bunches (Wilkinson *et al* 1994). For screw dislocations lying parallel to a free surface the requirement for no stresses acting to move the surface plane can be met by the use an image dislocation line. For edge dislocations however, surface tractions in addition to those from an image line must be used and the resulting corrections to the displacement fields have been given by Wilkinson and Hirsch (1995). These additional displacements change the strength of the ECCI contrast, and in some cases reverse its sense. This contrast reversal is of great importance since otherwise the dislocation bunches would be assigned the wrong character. The contrast observed for dislocation bunches in $Si_{1-x}Ge_x$ epilayers 1.0 and 0.17 μ thick were consistent with them acting to relieve the epilayer stress (Wilkinson *et al* 1994).

ECCI can often be used were TEM would be prohibited due to difficulties in preparing the necessary thin foils. One such example has been the examination of dislocation distributions in $Si_{1-x}Ge_x$ epilayer grown on a Si substrate that had been patterned with raised pads termed mesas. The relatively large height of the mesas 3 μm prevents plan view thin foils being prepared while cross sections intersecting the micron size features are also difficult to prepare. ECCI can reveal the dislocation distribution directly from the as grown structure (Wilkinson et al 1993b, Wilkinson 1994, 1996b). In square mesas dislocations were observed when the side length was 10 µm or larger, while on smaller mesas no dislocations could be seen. This could be due to either a lack of dislocation nuclei within the smaller structures, or a lack of stress to propagate them. Rectangular mesas were also examined in which it was found that no dislocation lines were propagated parallel to sides that were less than 10 µm apart (see fig. 18). The presence of dislocations across the width of these narrow mesas indicated that dislocation nuclei were available and so implies that there was insufficient stress to propagate them along the mesa length. This is consistent with the stress relaxation caused by the lack of lateral constraint at the mesa edges that was first analysed by Luyri and Suhir (1986). Further and complementary analysis of these mesa structures using EBSD will be presented in section IV D.

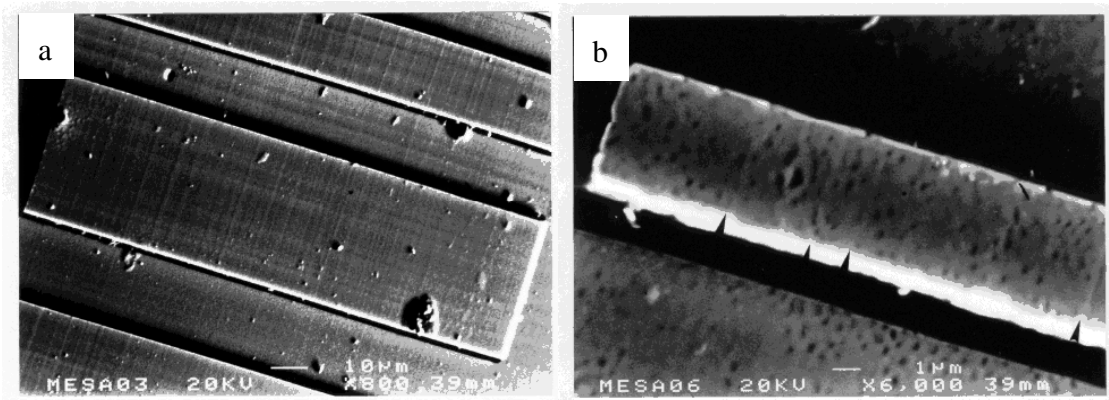

*Fig. 18     ECCI of misfit dislocations in $Si_{0.85}Ge_{0.15}$ epilayers grown on raised mesa pads patterned on a Si substrate. Dislocations run along both axes of the 40 µm wide mesa in (a), but only across the width of the 5 µm wide mesa in (b).*

*(b)  Observations of Fatigued Metals*

Another area of interest, in which research is currently being undertaken at Oxford is in using ECCI to image dislocation structures and strain fields in fatigued metals and alloys. This is not a new area for using electron channelling, since ECPs have been used by several groups in examining the plastic strain around fatigue cracks. Stickler *et al* (1971) used the degradation of ECP to map out iso-deformation contours around a creep rupture crack in a Discalloy specimen while Davidson and Lankford (1981) have examined plastic

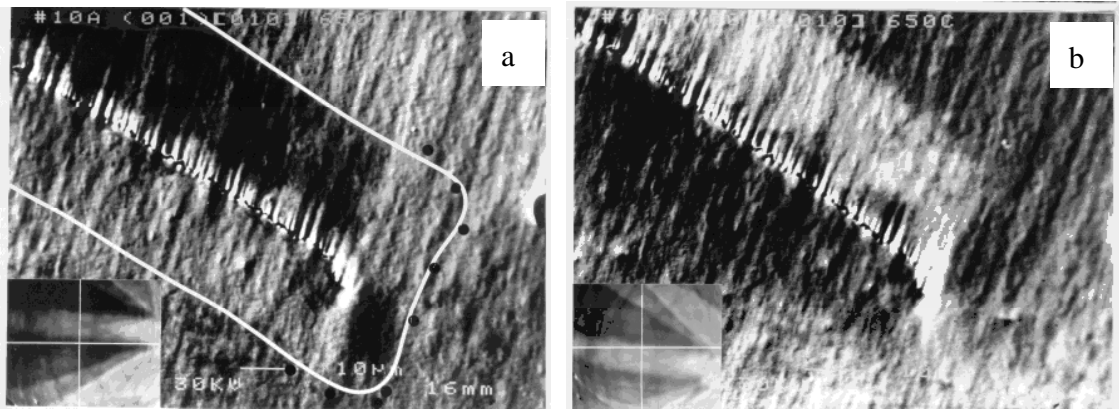

*Fig. 19    ECCI of a fatigue crack tip in a superalloy single crystal, in which the plastic zone is seen as dark and bright regions surrounding the crack. The diffraction conditions are shown in the inset ECPs and the image contrast inverts as **g** is reversed from (a) to (b). In (a) the edge of the plastic zone revealed by ECCI is drawn as a white line. (from Wilkinson, Henderson and Martin 1996)*

zones of fatigue cracks in Al alloy. More usually the technique has been limited to determining the size and shape of the plastic zone by using the disappearance from the ECP of a given high order channelling line to demark regions which have been plastically deformed. For example Tekin and Martin (1989) used the disappearance of the third order {111} line to find the size of plastic zones in a Ni-based superalloy for which the high strength can lead to plastic zones as small as 4 µm. Using ECP to map out plastic regions is a time consuming process and recently ECCI has been used to generate the same information (Wilkinson, Henderson and Martin 1996). Figure (19) shows the plastic zone of a mid-Paris regime fatigue crack in a single crystal superalloy. The dark circles in fig. (19a) denote points on the plastic zone boundary found using the ECP method, while the solid line shows the same boundary determined using ECCI. The results from the two approaches are in very close agreement but the ECCI method reveals the information more completely and is much quicker to use. The contrast is inverted when the sign of **g** is reversed as can be seen by comparing figs (19a) and (19b) which is strong evidence to confirm the crystallographic nature of the observed contrast. The sense of the contrast reverses across the crack plane and can be used to deduce the sense of the lattice plane tilting which is seen in fig. (19) to differ from the crack wake to ahead of the crack tip. Such qualitative information concerning the sense of the strain field gives a further advantage of ECCI over the use of ECP in this situation.

In addition to examining fatigue cracks, ECCI has also been used to examine the development of dislocation structures during cyclic deformation of un-notched specimens. Zauter *et al* (1992) have used BSEs in an SEM to reveal similar structures in bulk fatigued materials (mainly austenitic stainless steels) to those seen using TEM of thin foils. However, the diffraction conditions were not controlled in this work, but since polycrystalline materials were investigated some grains were in suitable orientations for the channelling contrast to be observed. Recently fatigued single crystals of Al and Cu have been examined using ECCI (Zhai *et al* 1996, Ahmed, Wilkinson and Roberts 1996) with the more striking results coming from the higher atomic number material. Initial observations were made on a Cu specimen prepared and kindly provided by Prof. Z. S. Basinski (McMaster University). This specimen had been cycled under a constant plastic strain amplitude of $\pm 2 \times 10^{-3}$ for $32 \times 10^3$ cycles which resulted in saturation of the cyclic shear stress and the formation of persistent slip bands (PSBs) in which the deformation is localised. The crystal was oriented with one face containing the primary Burgers vector so that it remained flat during the cycling and the shape of the PSB can be clearly seen at the sharp corner at its edge which had been prepared for that purpose using the method given by Basinski and Basinski (1984). ECCI obtained from the flat faces of such crystals are given in fig. (20). The first observations revealed bands running across the entire width of the specimen parallel to the trace of the primary slip plane, whose widths and positions corresponded exactly to the pattern of roughening seen on the orthogonal faces. The structure within these PSBs was found to consist of rotated cells measuring ~1 µm across as shown in fig. (20a). TEM micrographs of Cu after similar fatiguing show PSBs to consist of a series of regularly spaced walls of high dislocation density arranged in what is termed the ladder structure, while rotated cells revealed by ECCI are more typical of later stages of fatigue. Material was removed from the free surface by electropolishing and more observations made. ECCI then revealed the ladder structure that was initially expected (fig. 20b). The result shows the importance of the presence of a free surface on the development of the dislocation structure. Dislocation structures that occur prior to stress saturation have also been examined and fig. (21) shows that the elongated cell structure typical of the earlier stages of fatigue can also be revealed by ECCI. The advantage of ECCI over TEM in this work is that bulk specimens can be used for ECCI so that the evolution of dislocation structures leading to crack formation can be followed during interrupted fatigue tests on the same specimen.

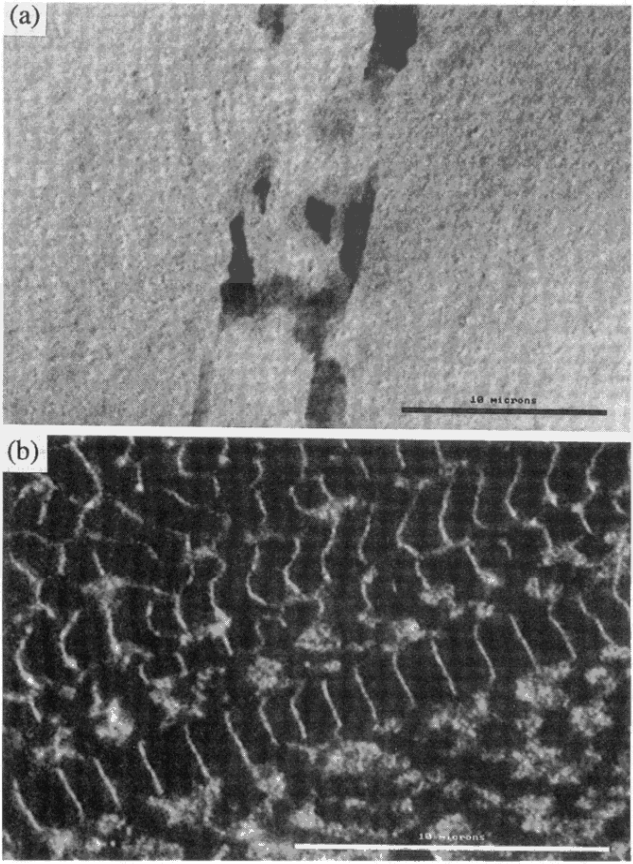

*Fig. 20*   *Structure seen within PSBs in a Cu single crystals fatigued beyond stress saturation. (a) rotated cells seen at the specimen surface, and (b) the ladder structure seen after electropolishing to reveal the interior of the specimen.*

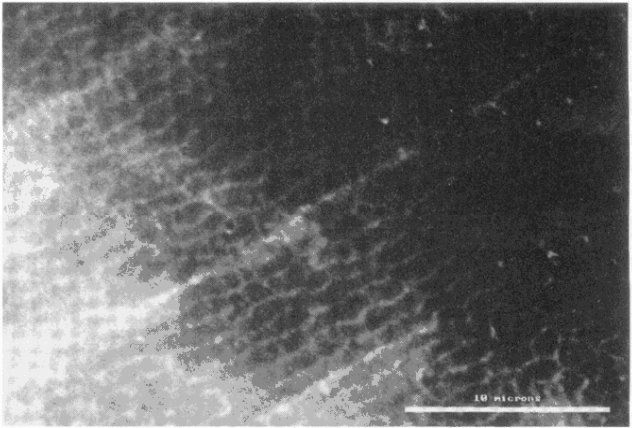

*Fig. 21    elongated cell structure seen in fatigued Cu single crystal prior to the formation of PSBs and stress saturation.*

### III C  Other SEM based diffraction contrast imaging techniques

There have been attempts to form diffraction images in the SEM using contrast occurring due to the channelling of electrons as they exit the crystal. Harland, Akhter and Venables (1981b) produced images of subgrains in Ni using what they termed a small angle detector (SAD) which was a scintillator type BSE detector subtending a capture angle of $8 \times 10^{-3}$ sterad at the specimen surface. The specimen was highly tilted (75°-80°) and the SAD centred at a takeoff angle of ~27°, again corresponding to the direction of the forward scattered peak in the electron distribution. Subgrains misoriented by small angles ($\leq 1°$) were clearly imaged and the contrast was ascribed to the fact that for the different subgrains a slightly different part of the EBSD pattern was collected by the SAD. Thus some areas were bright when a prominent Kikuchi band was brought onto the SAD, while others were darker since the Kikuchi band was shifted away from the SAD. The angular width of the SAD was 5.7° compared to the total angular width ($2\theta_B$) of ~2° of a Kikuchi band from a typical low order plane, so that one would expect the contrast to be rather less than that of a Kikuchi band. An increase in contrast would be expected on reducing the angular width of the SAD, though this would be at the expense at a reduction in signal level and probably result in an overall reduction in the signal to noise ratio. It is quite possible however that the contrast seen with the SAD is simply that due to the usual channelling of electrons as they enter the specimen. More recently Day (1993) has imaged subgrains and deformation around indents in a Ni-based superalloy using what amounts to an array of SADs. Three detectors are used each subtending an angular width of ~6° degrees at the specimen which is again tilted through a large angle (70°), as shown in fig. (22a). Images are obtained using each of these detectors for the same region of the specimen with the incident beam direction remaining constant. Occasionally subgrains show a contrast inversion when the images from two detectors are compared, an example of this is given in fig. (22 b, c). Day takes this as evidence that the contrast seen in the images is due to channelling of the exiting electrons forming an EBSD rather than the channelling of electrons entering the specimen responsible for contrast in ECCI. However, it is also possible that this effect occurs due to the asymmetries and inversions in conventional channelling contrast that are known to occur with the tilted specimen geometry when the position of the detector is altered, as was shown in fig. (8 and 12). Whatever the contrast mechanism Day combines the signals from the three SADs, one supplying a red signal, one green, and the last blue, to make a false colour image in which the subgrain structure is more clearly visible than in the any of the individual constituent images.

An important recent development has been the automation of orientation measurements from EBSD patterns, that will be discussed in section IV B. By measuring the orientation over an array of points an orientation image can be constructed in which grain and subgrain structure is revealed by changes in the local crystal orientation. In comparison to ECCI this orientation imaging is much more time consuming to carry out, though the images produced do contain quantitative data. Additionally the ECCI technique is sensitive to small lattice distortions and rotations that would go undetected in orientation images formed using EBSD. This was made clear by the recent study by Prior *et al* (1996) in which ECCI images formed using the tilted specimen-low takeoff angle detector geometry showed substructure within grains of geological specimens examined, which was not seen in EBSD examination of the same areas.

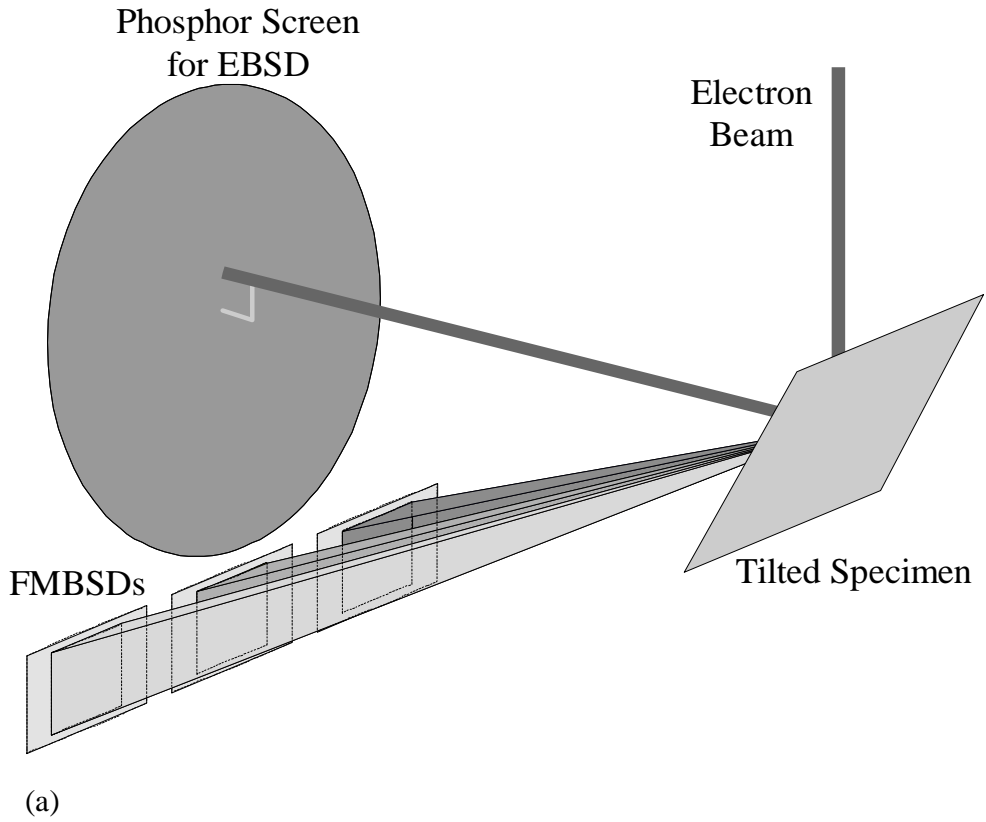

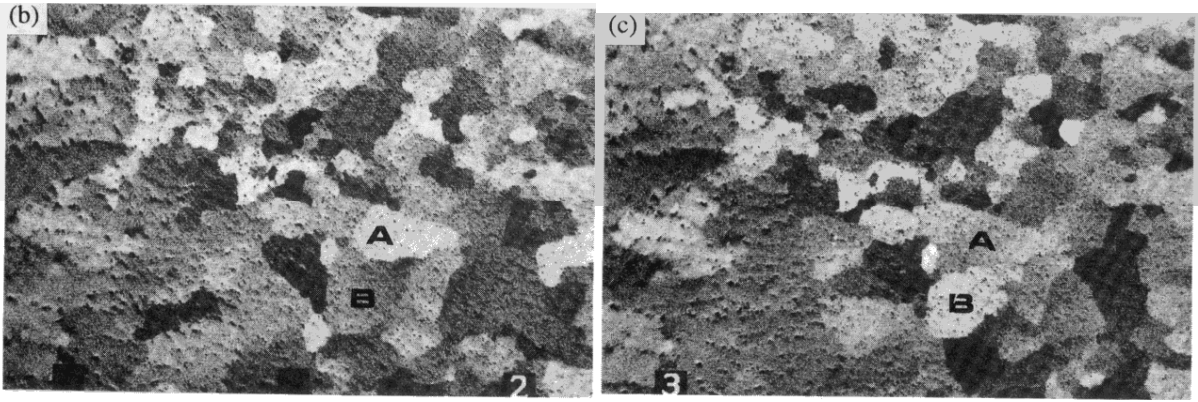

*Fig. 22* *(a) Geometry adopted by Day (1993) for an array of three BSE detectors held at low takeoff angle relative to a highly tilted specimen. Images in (b) and (c) were obtained from the same region of a Ni based superalloy using different detectors. The contrast between regions A and B is reversed from (b) to (c) although the specimen orientation is kept constant. (from Day 1993)*

# IV. ELECTRON BACK SCATTER DIFFRACTION

## IV A Formation of Electron Back Scatter Diffraction Patterns

An EBSD pattern is a form of Kikuchi pattern which is seen in the angular distribution of BSEs emitted from a crystal on which an incident electron beam is focused. The initial scattering events can be considered to produce a divergent source of electrons just below the specimen surface. These diverging fast electrons can be described in terms of Bloch waves which interact with the lattice as they propagate out of the crystal. Boundary conditions at the specimen surface again control the excitation of the different Bloch waves for a given exit direction in a similar manner to that for channelling of incident electrons entering the specimen. Reimer *et al* (1986) have given a good account of the reciprocity theorem that links ECP and EBSD. They show that theories very similar to those described in section II B for simulating ECP can also be used to calculate the form of intensity profiles across Kikuchi bands seen in EBSD patterns. However, in calculating the Bloch wave excitations the wave vector **-k** must replace that which would be used in simulating the ECP. The similarity in the forms of the two types of pattern is made clear in fig. (23) which shows an ECP and a similar region of an EBSD pattern centred on a <103> zone axis in Si obtained at 30 keV. The general forms of the two patterns are very similar, however the ECP shows considerably more detail than the EBSD pattern. The EBSD pattern was obtained with a large specimen to screen distance so as to achieve a higher angular resolution than is normal for the technique. Analysis of the recording system showed that the reduced clarity of the EBSD was not a simple result of the limited resolution of the pattern capture (Wilkinson 1996e). The difference in the sharpness of the two patterns is a real phenomenon and must result from the larger energy spread of exiting electrons involved in the interactions leading to EBSD, compared to the reasonably well defined energy of the incident beam electrons that produce the ECP.

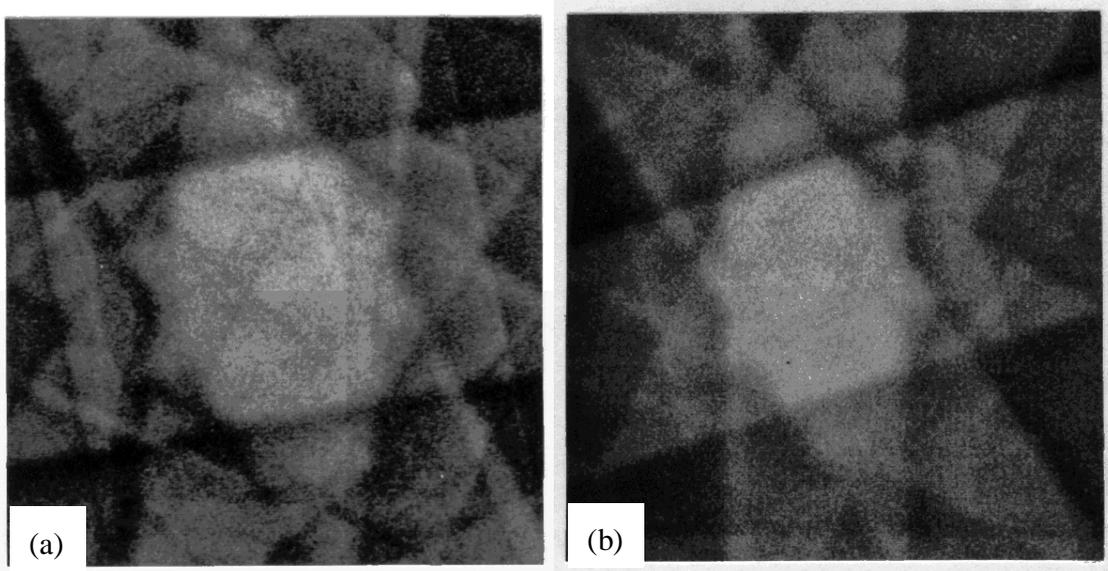

*Fig. 23*    *(a) An ECP from Si obtained at 30 keV and centred on a <103> zone axis, and (b) the corresponding region of an EBSD pattern. The ECP clearly shows much more detail.*

We turn now to the practicalities of recording EBSD patterns. Alam *et al* (1954) found that the patterns were much clearer when the incident beam was at a glancing angle to the specimen surface, though they noted that in some cases the patterns were even visible at normal incidence. A highly tilted specimen geometry (often 70.5° for calibration purposes) is now generally used for EBSD. The components of a typical modern EBSD system are shown schematically in fig. (24). Venables and Harland (1973) imaged the patterns using a scintillator screen mounted close to the tilted specimen inside the microscope's vacuum chamber which was viewed through a lead glass window. The patterns were recorded using a 35 mm film camera, and any measurements were made using prints from these negatives. The technique was much improved during the 1980s when Dingley and co-workers (Dingley 1984, 1989, Dingley et al 1987) used a low light level video camera to image the pattern on the scintillator screen in real time. Furthermore they integrated the signal from the camera with a graphical overlay from a computer so that measurements could conveniently be made from the patterns. This advance made the EBSD technique an important material science tool with which the orientations of large numbers of individual grains in a polycrystal could be determined reasonably rapidly. Further advances have been made through the 1990s to automate fully the pattern interrogation so that the orientations can be determined by computer directly from the patterns without the need for the intervention of an operator (Juul Jensen and Schmidt 1990, Wright and Adams 1992, Krieger Lassen *et al* 1992, and Adams *et al* 1992). This advance means that much larger data sets can be gathered, which should allow a better statistical representation of the texture and related properties to be obtained.

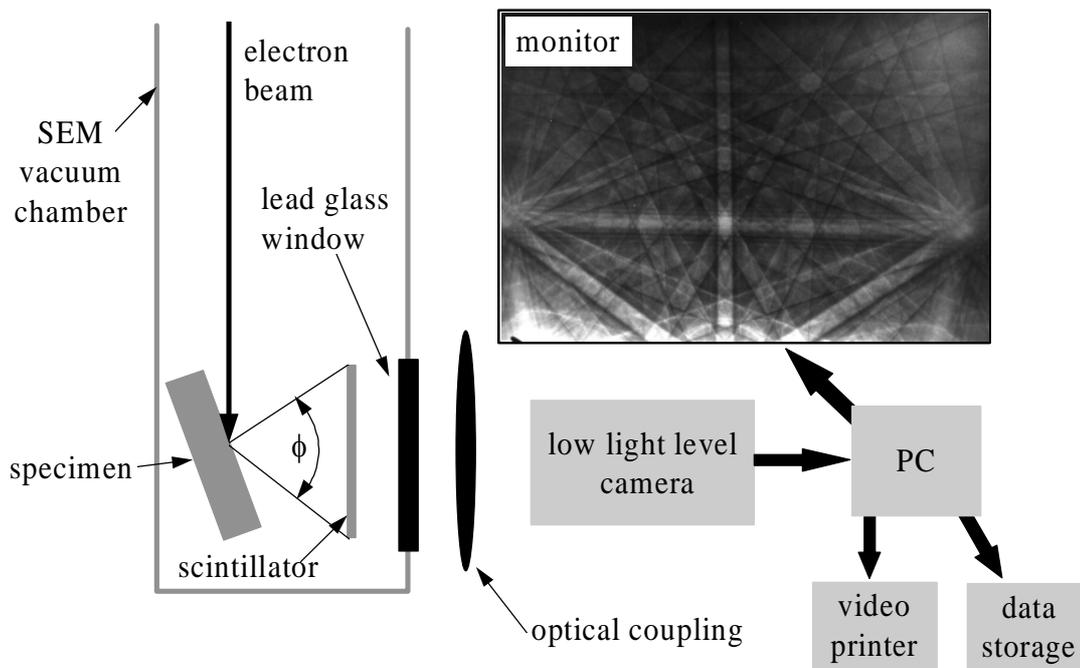

*Fig. 24      Schematic diagram showing the experimental set-up for recording EBSD and an example EBSD pattern.*

Compared to ECP the spatial resolution of EBSD is superior. This is because the pattern is formed using a stationary focused electron beam so that the incident beam diameter, determined by the lens aberrations and demagnification of the electron source, is generally sufficiently small that it does not limit the resolution. The spatial resolution is in fact determined by the spreading of the beam within the specimen prior to electron emission and so just as with conventional BSE compositional imaging the resolution varies strongly with specimen atomic number and the incident beam energy. Also because of the high specimen tilt used the beam spreading is quite anisotropic resulting in better spatial resolution parallel to the tilt axis than perpendicular to it. Surprisingly there have been few detailed investigations of the EBSD technique's resolution. This is probably because initial applications were to materials where the grain size ranged from 5 µm to 500 µm so that the spatial resolution presented no real limitations. Harland *et al* (1981a) quoted a spatial resolution of 200 nm perpendicular and 50 nm parallel to the tilt axis in their measurements from a Au specimen in a FEG SEM operated at 30 keV. In a subsequent paper Harland, Akhter and Venables (1981b) stated that subgrain boundaries in Au could be resolved at ≤80 nm by 20 nm though these values appear to refer to images obtained using a SAD (see section III C) in which the contrast may arise from electron channelling into the specimen rather than out of it as is the case in the EBSD technique. The resolution is, as expected, found to be markedly worse in lighter materials. Hjelen and Nes (1990) have given the most complete data set in their assessment of the resolution obtained in Al. Resolution was found to improve on reduction of the beam energy being 2.2 µm by 0.6 µm at 40 keV and 1.0 µm by 0.25 µm at 20 keV, while it was relatively insensitive to beam current. Decreasing the specimen tilt from 75° to 60° improved the resolution perpendicular to the tilt axis almost by a factor of two, while leaving that parallel to the tilt axis unaffected. The resolution of the EBSD technique, which is controlled by the beam-specimen interaction, is better than that obtained with ECP which is limited by instrumentation to at best ~2 µm for all materials. Spatial resolution has now become a more important issue for EBSD studies in important applications such as characterising Al interconnect tracks and the role of texture in controlling electromigration failures (e.g. Alvis, Dingley and Field 1995, Knorr and Rodbell 1996). It is clear that working at lower beam energies gives improved spatial resolution; however, the performance of detection systems currently employed become worse at lower energies and has tended to prevented much work being undertaken at energies below about 10 keV. Several groups are currently trying to improve EBSD detection systems to allow operation at lower beam energies (Barr and Brown 1995), with some even aiming at energy filtering the patterns (Mancuso *et al* 1994) which although experimentally difficult to achieve may eventually give improved resolution and contrast, and also possibly increase the level of detail present in the patterns.

### IV B  Orientation Determination

(a) *Measurement of Orientation*

Venables and bin-Jaya (1977) showed that EBSD patterns could be readily analysed to determine the orientation of the diffracting crystal. A major advantage in using EBSD for orientation method is that the patterns subtend a large angle so that it is usual for several major zone axes to be present in the same pattern, which is not the case with ECP which have much smaller angular ranges. The basic analysis method requires the crystallographic nature of at least two zone axes to be identified and their positions measured relative to the pattern centre, which is the closest point on the scintillator screen to the diffracting volume in the specimen. Calibration methods for determining the position of the pattern centre and the specimen to screen distance will be discussed shortly. Figure (25) shows the geometry of the orientation determination, in which the positions of two zone axes relative to the pattern centre are $(x_1,y_1)$ and $(x_2,y_2)$ and the zones have been identified as crystallographic directions $[u_1,v_1,w_1]$ and $[u_2,v_2,w_2]$ respectively. For this illustration we shall assume the crystal to be cubic, but it should be clear that the method is easily extended to other crystal types. A third direction $[u_3,v_3,w_3]$ perpendicular to the other two is first calculated from the cross product of $[u_1,v_1,w_1]$ and $[u_2,v_2,w_2]$, and the position $(x_3,y_3)$ on the screen of this direction is also found using the cross product of $(x_1,y_1,z)$ and $(x_2,y_2,z)$. To define the crystal's orientation we need to establish the crystallographic directions along some macroscopic reference axes in the specimen, which are generally taken to be three perpendicular edges of the specimen, **A**, **B**, and **C** in fig. (25). To calculate the crystallographic direction $[h_C,k_C,l_C]$ along the surface normal **C**, the angles $\gamma_1$, $\gamma_2$ and $\gamma_3$ between **C** and the directions $(x_1,y_1,z)$, $(x_2,y_2,z)$ and $(x_3,y_3,z)$ are measured. This requires accurate knowledge of the macroscopic alignment of the specimen relative to the EBSD detector so that the position where **C** intersects the screen is known. The scalar products, with **C**, of the crystallographic directions along the zone axes are thus related to these measured angles, so that $[h_C,k_C,l_C]$ can be determined from the simultaneous equations (2)

$$u_1 h_C + v_1 k_C + w_1 l_C = \cos(\gamma_1)\, (u_1^2 + v_1^2 + w_1^2)^{1/2}$$
$$u_2 h_C + v_2 k_C + w_2 l_C = \cos(\gamma_2)\, (u_2^2 + v_2^2 + w_2^2)^{1/2} \qquad (2)$$

$$u_3\,h_C + v_3\,k_C + w_3\,l_C = \cos(\gamma_3)\,(u_3^2 + v_3^2 + w_3^2)^{1/2}$$

Similar equations can be written and solved to find the remaining directions $[h_A,k_A,l_A]$ and $[h_B,k_B,l_B]$ corresponding to **A** and **B**. Having calculated the crystal orientation the positions on the screen of prominent Kikuchi bands and zone axes can be calculated and displayed over the original pattern. This allows the user to verify that the orientation calculated is indeed correct.

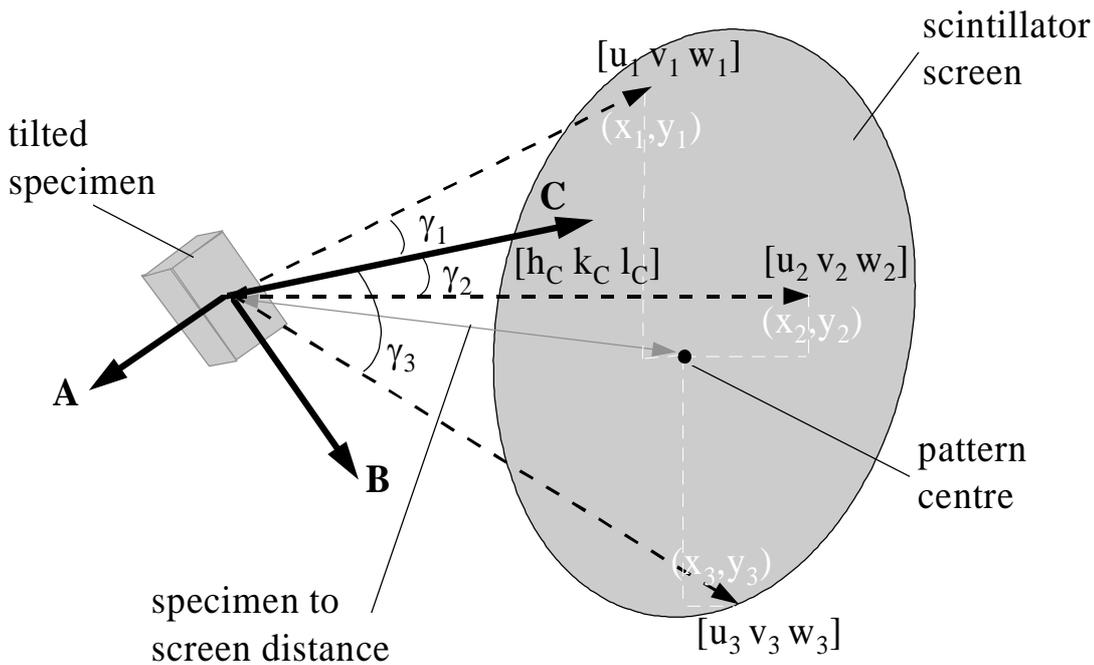

*Fig. 25    Geometry used to determine a crystal orientation from the observed EBSD pattern.*

Subsequent developments led to an analysis that required the user simply to indicate the positions of three zone axes, the crystallography of which need not be indexed by the user (Dingley 1989). The angles between these directions were calculated and compared to a table of known angles between prominent zone axes in the crystal system being examined. The best fit of the experimental angles to the known ones is used to establish the nature of the three zone axes and the analysis then proceeds as described above. Occasionally this method incorrectly identifies the zone axes leading to the wrong orientation which is made clear when the calculated EBSD pattern is compared with the real one. When this happens the user rejects the result and the next best fit of interzonal angles to those measured is used to characterise the zones. Generally well over 90% of patterns are correctly analysed in the first attempt, when only three zones are used in the analysis. If further zones are located the success rate of identifying them can be improved, and a least squares method used to improve the orientation determination.

It is clearly desirable for the orientation measurement to be performed completely by a computer and several groups have examined methods to enable the computer to recognise the positions of zone axes or Kikuchi bands in EBSD patterns (Juul Jensen and Schmidt 1990, Wright and Adams 1992, Krieger Lassen *et al* 1992, and Adams *et al* 1992). The method that most workers have now settled on involves forming a Hough transform of the pattern so that the linear bands are transformed into spot or peak-like features in the transform which are easier to locate. Typically the first step in the image processing is to remove any background intensity variation from the raw EBSD pattern, and then rescale the intensities to fill the available grey levels. The angular resolution of the pattern is also generally reduced so that the image consists of about 100 by 100 pixels. This acts to increase the speed of the analysis by reducing the number of pixels and also affords some integration of intensities over local blocks of pixels which improves the signal to noise ratio. The Hough transform of this image is then calculated. In this transformation the intensity, $I(x, y)$, at each pixel in the EBSD image is added to the value of pixels $I(\rho,\theta)$ in the transform, which are all initially set at zero, where $\rho$ and $\theta$ are the Hough parameters of all lines through the point (x,y) so that

$$\rho = x \cos(\theta) + y \sin(\theta) \qquad (3)$$

Figure (26) shows that the Hough parameters give the perpendicular distance $\rho$ from the origin to a line through the point (x,y) whose normal is inclined by an angle $\theta$ to the x axis. Each point in the original image then corresponds to a sinusoidal curve in the transform, while a line in the original corresponds to a single point in the transform. A bright line one pixel wide in the image would thus become a single bright point in the Hough transform which could then be found by very standard peak locating methods. In the EBSD patterns the finite width of the bright Kikuchi lines produces peaks of finite extent in the transform. These peaks have a butterfly-like shape, and rather than simply search the transform for the largest peaks it is first filtered, using a kernel designed to output high intensities wherever the butterfly motif is located in the transform. The size of the butterfly shape is dependent on the width of the Kikuchi bands, so that different sized filters are used as the beam energy is altered. Once the peaks are located in the filtered Hough transform this gives the Hough parameters defining the locations of the prominent Kikuchi bands in the EBSD pattern. The methods described previously can then be used first to determine the nature of the Kikuchi bands in this case from comparison with known interplanar angles, and then to calculate the orientation. The method is reasonably reliable and commercial software is now available to undertake such analysis which allows the orientation to be established in a few seconds. By allowing the computer to also control the position of the probe on the specimen many thousands of measurements can be obtained so that much fuller data sets and images can be obtained compared to what could be obtained using the manual indexing methods.

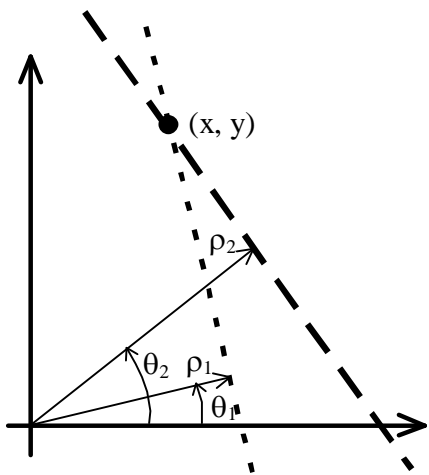

Fig. 26    Hough parameters $\rho_i, \theta_i$ defining lines through the point (x,y).

*(b) Calibration*

Accurate determination of the pattern centre position and the specimen to screen distance are crucial in determining the reliability of the final orientation measurement. The most commonly employed method uses a crystal of known orientation. Typically the crystal is a cleaved piece from a Si wafer which has a surface plane of (001) with sharp straight edges along the [110] and [$\bar{1}$10] directions. The crystal is tilted through 70.53° away from normal incidence and towards the scintillator screen. At this angle the [114] becomes normal to the incident beam and strikes the screen at a right angle. The [114] zone axis then marks the position of the pattern centre. The specimen to screen distance is then found using simply trigonometry and the measured distance from [114] of another zone axis, which makes a known angle to [114]. The position of the pattern centre moves up and down the screen if the specimen is moved such that the interaction volume moves along the optic axis of the microscope. This shift in the pattern centre is calibrated against by recording the shift in the position of the [114] zone axis as a function of the working distance of the microscope, which is determined from the objective lens excitation required to bring the specimen surface into focus.

One problem with this method is that the effects of incorrect alignment of the calibration specimen are not always seen in the pattern, and such misalignment leads to error mainly in the pattern centre location, which affects all subsequent measurements. Day (1993) has suggested using the geometry of the kite shape formed by the [114], [101], [111] and [011] zone axes around the [112] zone axis to detect any misalignment of the calibration specimen. This can be detected through the observed ratios between the distances from these zones to [112], and the orientation of the calibrant crystal adjusted. This ensures that the [114] zone axis marks the correct position of the pattern centre, which Day found could be up to 4° away from the initial alignment. Day estimates that having used the kite method for calibration absolute and relative grain orientations can then be determined at an angular resolution of ±0.7° and ±0.4° respectively, which offers improvement over the ±2° and ±1° resolution that can be obtained with the standard crystal, or known orientation method.

A further improvement in the accuracy of pattern centre location has been given by Day (1993). In this method two wires are mounted parallel to the face of the scintillator screen, one being quite close to the screen and the other at some distance, so that they define a plane perpendicular to the screen. Generally the two wires cast two shadows onto the screen, however; when the diffracting volume on the specimen lies in the plane defined by the wires, the two shadows coincide and only a single shadow is seen. A second similar set of wires are placed in front of the screen defining another plane perpendicular to the screen, so that when both pairs of wires cast single shadows the pattern centre is located by the intersection of the shadow lines. When the shadows are split into pairs the pattern centre can still be determined providing the distances between the wires and the screen are known. Using such crosswires Day (1993) can locate the pattern centre to ±50 µm, resulting in absolute and relative orientation measurements good to ±0.5° and ±0.2° respectively.

*(c) Applications*

The number of texture studies using the EBSD technique is growing rapidly, so that it is impossible to give a comprehensive review here. The examples given here were selected to try and demonstrate the advantages of EBSD over the more usual x-ray and neutron diffraction methods for analysing texture.

One major difference between EBSD and x-ray and neutron diffraction is that EBSD makes measurements on a grain by grain basis while the others sample a volume containing many grains. An important question that arises is: how many grains must be sampled with EBSD to obtain a good representation of the overall texture? This problem has been addressed by Baudin and Penelle (1993) who compared orientation distribution functions (ODFs) constructed from x-ray pole figures, with those constructed from EBSD measurements from a recrystallised Fe-3% Si sheet. In this case there were two main components to the texture; {111}<112> increasing from the centre of the sheet and {100}<012> decreasing from the centre of the sheet. For the EBSD results to find the principal texture components only ~100 measurements needed to be made, but in order to get good accuracy concerning the strengths of the texture components ~1000 grains needed to be sampled. This example shows that to attempt a representation of the texture with statistics as good as those generated by the volume averaged diffraction methods, the size of the data sets required is sufficiently large to make a fully automated EBSD system a requirement for routine work of this kind. With such a system performing a measurement every few seconds, the ODF can be established in much the same time scale as is typical using x-rays or neutrons to generate 3 pole figures in order to calculate the ODF.

A further difference between x-ray or neutron diffraction methods and EBSD is that EBSD measures the complete 3 dimensional orientation of each crystal directly, while the volume averaged techniques measure only the orientations of the selected diffraction planes (2 dimensional information) and give no information about the crystallographic directions within that plane. In constructing the 3 dimensional ODF from x-ray or neutron data at least two pole figures are required, and despite 20 years of work the methods for combining the pole figure information to form the ODF is still under discussion. The total ODF consists of both odd and even parts but in the analysis of x-ray and neutron data only the even part can be found directly. The EBSD measurements allow both odd and even parts to be obtained directly and the measurements of Baudin and Penelle (1993) have shown that the odd part can account for 20% of the total ODF.

The fully automated EBSD systems now give direct competition to x-ray diffraction systems for analysis of macroscopic texture. However, the spatially specific nature of the EBSD measurement allows analysis that cannot be undertaken using the macroscopic x-ray and neutron techniques. With EBSD the orientations of key features of the microstructure can be assessed individually. An example taken from work by Miodownik *et al* (1996) concerns the secondary recrystallisation of an oxide dispersion strengthened Ni-based alloy (MA754). The microstructure of an extruded bar consisted mainly of fine 400 nm diameter equiaxed grains, which exhibited a weak <001>-<111> fibre texture, shown in fig. (27a). Observation of etched specimens in the SEM revealed the presence, at a low volume fraction, of a few larger grains, that were often elongated, within the equiaxed matrix (fig. 27b). In the SEM it was possible to locate several of these grains, possessing significant size advantages, and being likely candidates for recrystallisation embryos. The orientations of approximately 50 such grains were measured and are shown in fig. (27c), where a much more pronounced texture is evident compared to the matrix. It is important to note that x-ray or neutron diffraction would not allow such measurements to be made, and because of the low volume fraction of these large grains TEM analysis could not be used due to the prohibitively large number of foils that would be required. The orientations measured for the large grains includes the <001> fibre texture that is found to be very strong in the final fully recrystallised state.  Selection of <001> oriented grains from other potential recrystallisation embryos is thought to depend on mesotexture at some of the <001> oriented embryos giving reduced solute pinning so allowing their growth to be triggered prior to other grains (Miodownik *et al* 1996).

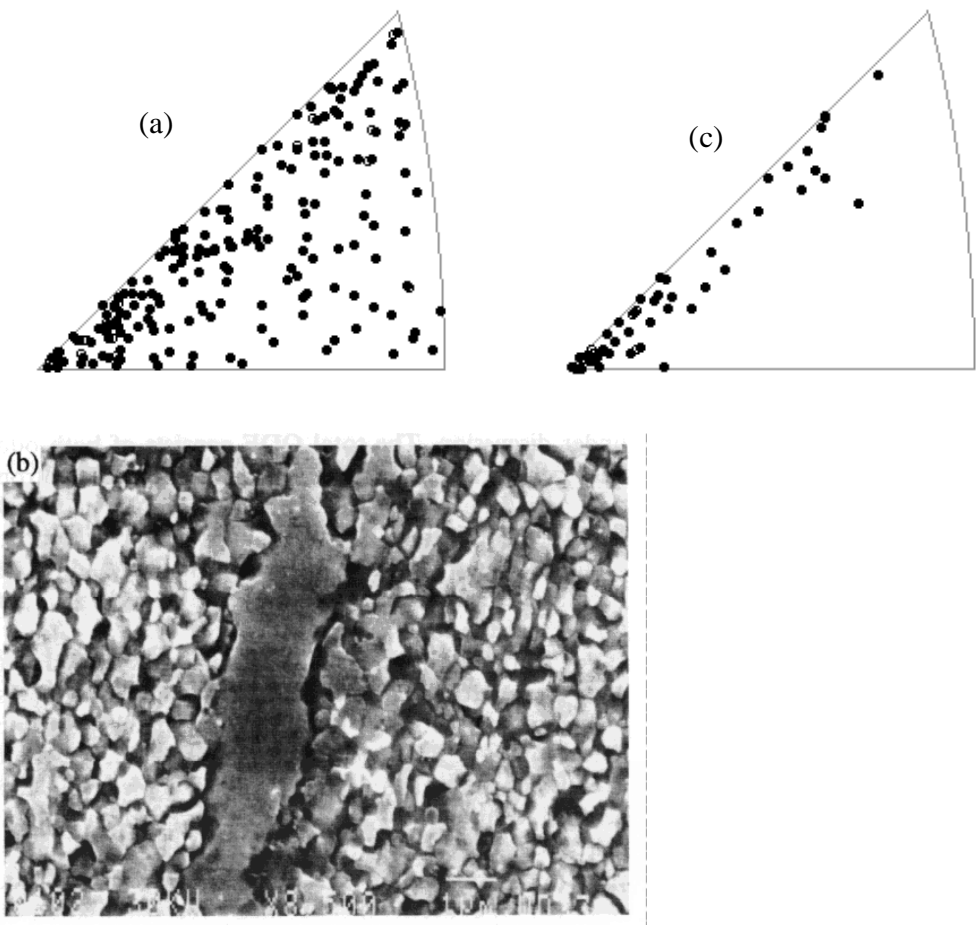

*Fig. 27* *(a) Inverse pole figure showing orientations obtained from the initial fine grained structure of a Ni based oxide dispersion strengthened alloy (MA754). (b) An example of a grain with a markedly larger size that were found in the alloy at very low volume fraction. The orientations of these larger grains is shown in (c).*

Since the orientations of neighbouring grains can be determined using EBSD it follows that misorientation between the two grains can also be readily calculated (see for example Randle 1993).  This has led to use of the EBSD technique for determining the distributions of grain boundary types in polycrystals, which has been termed mesotexture.  Much of the work on mesotexture has concentrated on characterising the evolution of populations of different grain boundary classes (low angle, high angle, coincident site lattice - CSL boundaries) during grain growth.  Randle and co-workers (Randle and Brown 1988, 1989, Randle 1990, 1991, Furley and Randle 1991) have made extensive studies of mesotexture development in Ni and austenitic steel.  Histograms were produced by Furley and Randle (1991) to show the populations of different boundary classes in a Ni specimen with only a weak macrotexture, after an initial treatment giving a 50-250 µm grain size, and then after grain growth to a 100-250 µm grain size during an anneal at 900°C.  After the anneal the $\Sigma = 3$ and $\Sigma = 9$ boundaries increased in frequency, and were significantly more frequent than would be expected from a random selection of grain pairs in a completely untextured material.  Larger data sets are now becoming available, and more sophisticated representations of the mesotexture such as the orientation coherence function are being developed.  Wang, Morris and Adams (1990a,b) have examined the orientation and its spatial distribution in commercial purity 1100 Al in both an as-cast ingot, and after a moderate 20% channel die compression.  The as-cast material showed relatively weak texture, and no spatial coherence in the texture distribution, but after the compression strong evidence was found that grains oriented at Euler angles $(\phi_1, \Phi, \phi_2)$ where preferentially adjacent to grains oriented at $(\pi+\phi_1, \Phi, \phi_2)$.  In particular orientation coherence functions were shown for grains with the orientation (90°, 25°, 50°) which showed a marked tendency for neighbouring grains along the normal and rolling direction to have the orientation (270°, 25°, 50°). Such results show that grain-grain interactions during deformation result in texture that does not develop in a spatially uniform manner.

The orientation relationship accounts for three of the five degrees of freedom at a grain boundary and the other two are required to define the grain boundary plane. These two can be determined by combining the EBSD orientation measurement with the geometry of plane inclination found either by preparing a sharp corner (Randle and Dingley 1989) or by serial sectioning to remove a known thickness of material (Randle 1995a,b). Randle's study in nickel showed CSL boundaries other than $\Sigma = 3$ tended to be either high indexed asymmetrical tilt boundaries or have irrational boundary planes. Of the $\Sigma = 3$ boundaries about half were found to be asymmetrical tilt boundaries that were displaced from the symmetrical 111/111 coherent twin geometry, while the 211/211 incoherent twin was not observed due to its higher energy.

EBSD gives the ability to select features of the microstructure for orientation analysis, and to probe grain boundary misorientations and boundary plane types. Such analysis can not be undertaken using the macroscopic x-ray or neutron diffraction techniques, and comparison with TEM based methods is favourable since with EBSD the measurements can be made from bulk specimens, that are easily prepared, handled, and oriented to external references axes such as extrusion, or rolling directions.

### IV C  Phase Identification

As with all diffraction the symmetry in the crystal is reflected in the EBSD pattern. The extent to which the different symmetry groups can be distinguished from first principles using EBSD has been investigated (Baba-Kishi and Dingley 1989a,b , Dingley and Baba-Kishi 1990). The analysis starts by locating the prominent zone axes and identifying which of 10 Laue groups (6mm, 4mm, 3m, 2mm, m, 6, 4, 3, 2, 1) describes the symmetry at the zone, first by a visual assessment and then by more careful measurement. It is necessary to obtain several EBSD patterns covering different overlapping parts of the total pattern so that the inter-relation of sufficient Laue groups can be assessed. The large capture angle of the EBSD technique thus makes it much more preferable than ECP for identification of crystal types. The Laue groups of the individual directions are then combined to assess the possible crystal point group. Of the 32 point groups it appears in principle to be possible to distinguish unambiguously 23 of them. The remaining 9 groups can be placed into 4 classes, and if patterns can be obtained along directions 180° apart then it is possible to split these groups further. There have been attempts to take this analysis further and try to distinguish space groups (Baba-Kishi and Dingley 1989b). Having established the point group symmetry, the observed Kikuchi lines are recorded and systematic absences noted. For this type of analysis to have any chance of success it is imperative that high quality patterns are obtained and the reflections present in many systematic rows must then be assessed. In his analysis of EBSD from Si, Baba-Kishi (1990) observed that of the hk0 reflections 620, 260 and 420 were definitely present while 310,130 and 210 were definitely absent. From this the conclusion was drawn that the cause of the observed absences was a d-glide in the (001) plane, however, although this is consistent with the observations other possibilities exist, in this case face centring is also consistent with the data. This example illustrates quite clearly that such analysis should be limited to finding consistency with a possible space group rather than trying to identify the group outright. Further complications to such analysis include anomalous absences caused by dynamical diffraction effects and possibly the effects of high densities of stacking faults.

A more practical approach to phase analysis is to establish first the elements present in a phase and their approximate proportions using x-ray micro-analysis, and use this to establish a list of possible crystal types. Kikuchi maps can be readily simulated for each of these phases and compared to those obtained experimentally. Randle and Laird (1993) have used this approach to distinguish and examine the orientation relations between hexagonal $M_7C_3$, and orthorhombic $M_3C$ eutectic carbides in white cast irons. Dingley, Baba-Kishi and Randle (1995) have produced an atlas of EBSD patterns from a selection of materials so as to give indexed examples of patterns from the different symmetry types. This approach is in line with common practice in TEM where chemical and diffraction data are combined in phase analysis problems, and for the future integration of micro-analytical and diffraction capabilities on the SEM is clearly an attractive prospect.

### IV D  Strain Measurement

Neutron and in particular x-ray diffraction techniques are quite advanced and allow for accurate measurements of lattice parameters and strain. However, in many important areas of material science behaviour is controlled by the local distributions of stress and strain rather than global averages. This has been a strong driving force in the development of methods to probe both elastic and plastic strains with the good spatial resolution afforded by the EBSD technique.

*(a) plastic strains*

As was discussed for ECP in section II C, the distortion of the lattice planes that arises when dislocation are introduced into the crystal results in a degradation of EBSD patterns. This effect is demonstrated in fig. (28) and forms the basis of methods to measure the local plastic strain. Work in this area has taken the engineering approach of first establishing the strain variation of some parameter quantifying the pattern quality using a set of calibration specimens deformed by known amounts, and then making measurements on the components of interest. Quested, Henderson, and McLean (1988) were the first to use EBSD for plastic strain measurements. They found that when the patterns were obtained while the beam was scanned over the specimen the pattern quality improved as the magnification of the SEM was increased until at some critical magnification it became the same as that obtained with a stationary focused beam. They found that this critical magnification increased as the deformation in the specimen increased. In their method the quality of the patterns were simply compared by eye and so the measurements were somewhat subjective. Subsequent efforts attempted to remove this subjectivity by using image analysis in particular Fourier analysis to generate objective parameters quantifying the pattern quality (MacKenzie and Dingley 1986, Wilkinson and Dingley 1991, 1992, Wilkinson, Gonzalez and Dingley 1992, Krieger Lassen *et al* 1994, and Mukherjee *et al* 1995). The image quality parameters constructed all make use of the fact that the sharper patterns from unstrained material contain higher frequency intensity variations within them compared to the rather blurred patterns from strained material. Wilkinson and Dingley (1991) showed that such methods could be used to distinguish local plastic strains at a sensitivity of $\sim\pm 1\%$, in Al 6061 alloy deformed in uniaxial tension. The EBSD technique has been used to examine the distribution of plastic deformation in Al matrix composites reinforced with SiC fibres (Wilkinson and Dingley 1992, Wilkinson, Gonzalez and Dingley 1992 and Mukherjee *et al* 1995), while Krieger Lassen *et al* (1994) used the EBSD pattern quality to distinguish between deformed and recrystallised regions of Al and Cu specimens. It should be noted that the image quality is really a function of the lattice rotation occurring within the diffracting volume which results from the dislocations, in particular the excess of dislocations of one sign, generated as a consequence of the plastic deformation.

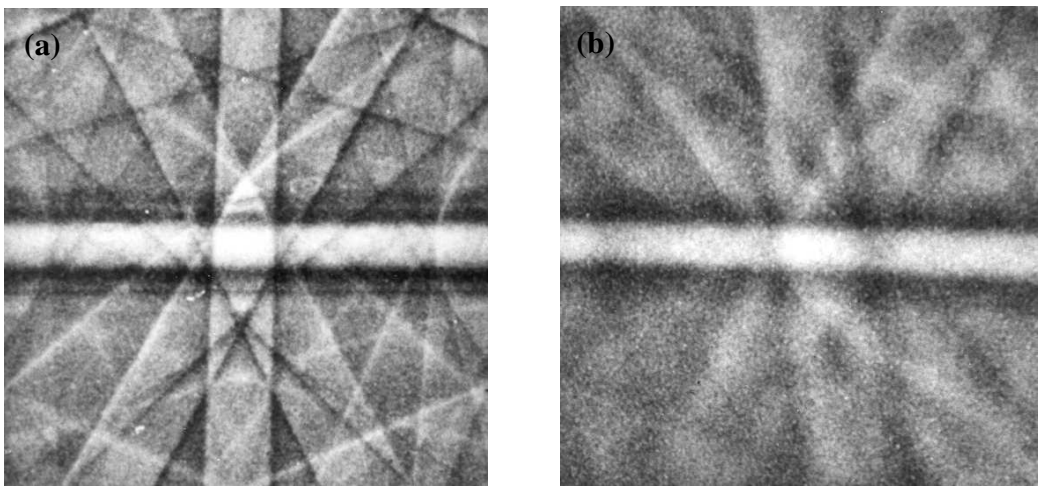

*Fig. 28    Two EBSD patterns from Al 6061 alloy, (a) from a annealed specimen, and (b) from a specimen deformed by 8.5% strain. The pattern from the deformed material is markedly more diffuse.*

An important assumption in such studies is that similar deformation processes, leading to similar dislocation structures, occur in the calibration specimens as in the test specimen, which may not always be the case. Wilkinson (1991) noted that often one Kikuchi band appeared more diffuse than another crystallographically similar one within the same EBSD pattern. By comparing the relative diffuseness of several bands in the same pattern, a correlation was established between the degree of diffuseness and the magnitude of **g.<b>**, consistent with local bending due to a local excess of dislocations with Burgers vector **<b>**.

An alternative approach is to measure the local rotations directly so as to assess the local excess density of dislocations of one sign. Randle, Hansen and Juul Jensen (1996) have given an example of this approach in which the increased lattice rotation in the region of grain boundaries and triple junctions in polycrystalline Al deformed by rolling was analysed. Figure (29) shows an example where lattice rotations were measured over an array of points to form an image of the region close to a triple junction in Al after a 5% reduction. In this case the increased dislocation activity close to the triple junction appears to have occurred predominantly in only one of the grains. Such an approach has the attraction that the measured orientations can potentially be linked quantitatively to the local excess densities of dislocations of one sign which cause the observed lattice curvature.

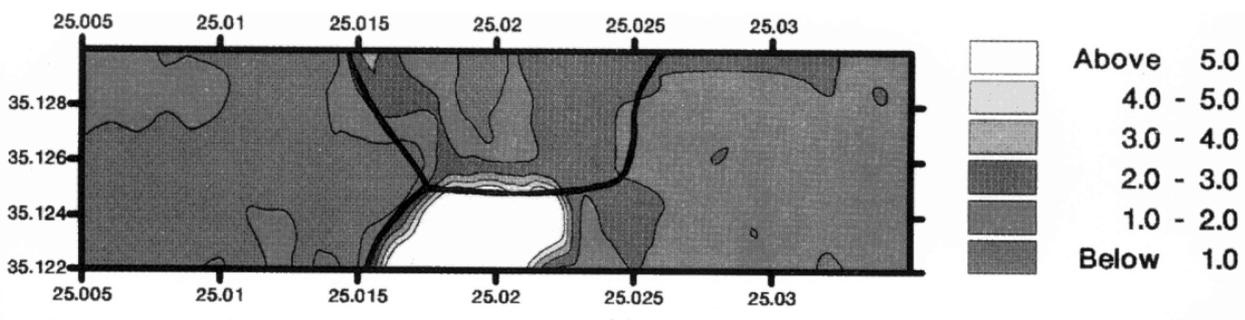

*Fig. 29*   *Misorientation distribution close to a triple junction in 5% deformed Al. The misorientation is given in terms of the parameter*
$D_E = [(\phi_1 - \langle\phi_1\rangle)^2 + (\Phi - \langle\Phi\rangle)^2 + (\phi_2 - \langle\phi_2\rangle)^2]^{1/2}$, *where $\phi_1$, $\Phi$, and $\phi_2$ are Euler angles describing the local orientation and $\langle\phi_1\rangle$, $\langle\Phi\rangle$, and $\langle\phi_2\rangle$ describes the reference orientation measured at the centre of each grain. (from Randle et al 1996)*

*(b) elastic strains*

There are two possible approaches to determining elastic strains from EBSD, the first uses shifts in Kikuchi line positions and the second shifts in zone axis positions. The position of Kikuchi lines in EBSD (and channelling lines in ECP) depends on the crystal orientation and the Bragg angle for the diffracting planes. The line positions are thus functions of the lattice plane spacing which is of course strain dependent and the electron wavelength. On differentiating the Bragg equation it is clear that strain should produce larger shifts in the positions of high order lines, and this coupled with their narrow width makes them beneficial to strain sensitivity. Wilkinson (1996e) has estimated that to measure an elastic strain of 0.1% it is necessary to locate the positions of lines of the order of {880}. These lines (and even higher orders) are easily seen in ECP, but the lack of detail in EBSD patterns (see fig. 23) means that they are not typically to be seen using EBSD. More typically lines such as {660} or even {440} represent the highest order lines that can be seen in EBSD patterns and these generate strain sensitivities approximated to be 0.3% and 1.0% respectively. Clearly this approach does not have sufficient sensitivity to make useful elastic strain measurements.

An applied elastic strain can also change the angles between zone axes. Troost *et al* (1993) developed image processing methods to measure the shift in zone axis position caused by tetragonal distortion of a $Si_{1-x}Ge_x$ epilayer grown on Si, and similar methods were developed independently by Wilkinson (1994) in measuring small lattice rotations in epilayers grown on substrates patterned with mesas. Subsequent work by Wilkinson (1996b) showed that the method was sensitive to elastic strains of 0.02% and rotations of 0.01°. Troost *et al* (1993) and Wilkinson (1994, 1996b) both obtained the EBSD patterns with the specimen to screen distance increased from the usual ~30 mm for orientation measurement, to ~80 mm and ~140 mm in the respective works. This improves the angular resolution of the EBSD patterns and also decreases the sensitivity of the measurements to movement of the diffracting volume at the specimen.

For the strain measurements patterns were recorded as pairs one from the unstrained substrate and one from the strained epilayer. Troost *et al* (1993) made measurements of the angular shift at the [111] zone axis, while Wilkinson (1996b) measured the shifts ($\alpha$) at several zone axes inclined by different angles $\theta$ from the [001] surface normal. The zone axis shifts were determined from a cross-correlation of the pattern from the epilayer with that from the substrate. Wilkinson (1994, 1996b) made use of high pass filtration prior to calculating the cross-correlation function, so that the peak in this function was made sharper, because of the removal of the low frequency variations, and hence was easier to locate. The measured values of $\alpha$ agreed well with their expected variation with $\theta$, as can be seen in Fig. (30) showing data from 4 epilayers with different Ge contents.

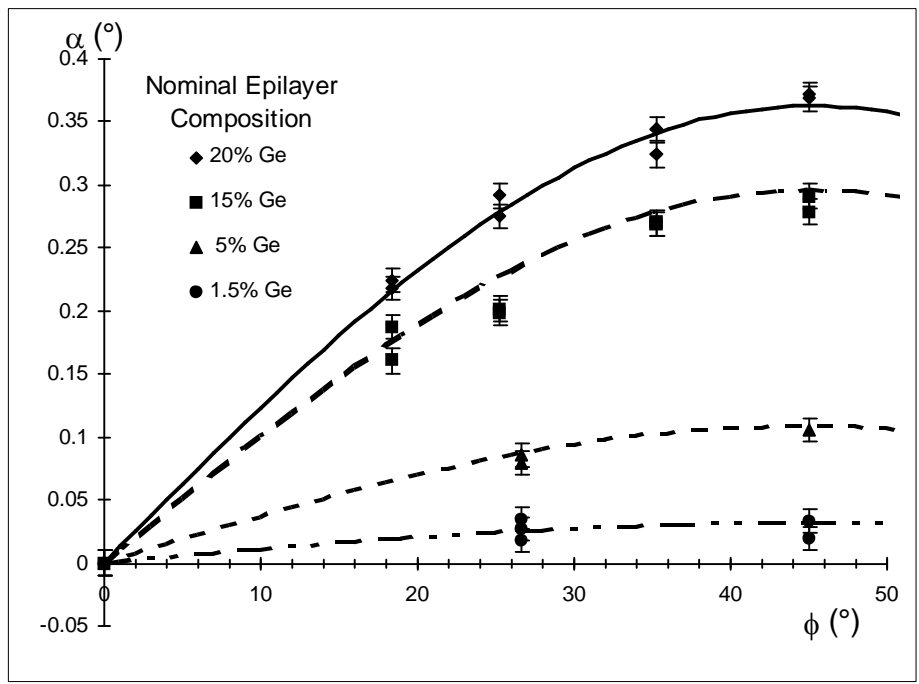

*Fig. 30   Shifts ($\alpha$) in the direction of various zone axes in EBSD from tetragonally distorted $Si_{1-x}Ge_x$ epilayers relative to patterns from their Si substrates.  The shifts show the expected variation with inclination ($\phi$) of the zone axis from the surface normal, and epilayer composition.*

The measured strains were also in good agreement with high resolution x-ray diffraction data.  This method can in principle be used to determine any strain that changes the shape of the crystal, but is insensitive to a pure dilatation or contraction.  A further condition for its use is that absolute strains can only be given if comparison can be made with patterns from a region of crystal at a known strain.

Wilkinson also examined the rotation of various zone axes across the width of $Si_{1-x}Ge_x$/Si mesa structures that were also examined using ECCI.  In wide rectangular mesas lattice rotations were limited to the outer edges of the structures where the lack of lateral constraint allows the epilayer to bow out to relieve the misfit strain.  In narrower mesas, in which ECCI indicated that stress driving the generation of misfit dislocations had been relaxed, the rotations spread across the entire mesa width.  The EBSD measurements corresponding to the structures observed using ECCI in fig. (18) are given in fig. (31).  These EBSD measurements confirm that the mesa structures can relax elastically, and complement the ECCI observations well.

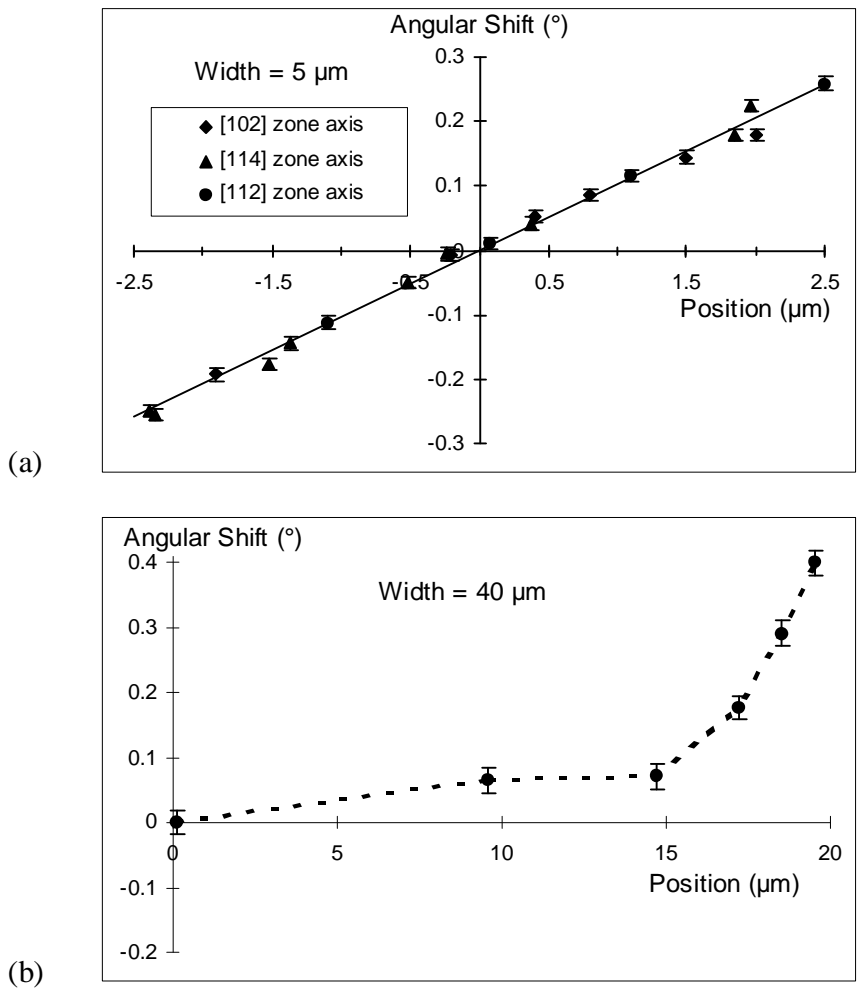

*Fig. 31  Angular shifts in zone axes directions measured as a function of position across the width of $Si_{1-x}Ge_x/Si$ mesas. (a) for 3 different zone axes and a narrow 5 µm wide mesa and (b) for a <102> zone axis and a 40 µm wide mesa.*

## V. CONCLUSIONS

It is clear that techniques using back scattered electrons from bulk materials, whose intensities are controlled by diffraction, provide the material scientist with a powerful addition to the armoury available for materials characterisation. The ECP, ECCI and EBSD techniques all have their own special advantages and applications, and are to some extent complementary. Thus, the ECCI technique is likely to have increasing applications in the imaging of regions of large plastic deformation, (e. g. around cracks) where the orientation dependent contrast can give information on the nature of the plastic strains. The important advantage over TEM studies is of course that bulk materials can be investigated. Again, the ECCI technique can be used to image second phase particles, and to study their spatial distribution. The EBSD technique has wide applications in the determination of local crystal orientation, and relative orientations across grain boundaries, with high spatial resolution. The additional spatial information on textures provides an important advantage compared with conventional x-ray or neutron diffraction techniques. With the automated techniques of pattern analysis, textures with good statistics can be obtained.

The EBSD technique is also providing a tool for measuring strains on a scale of <1 µm, on the surface of bulk materials; information which is not possible to obtain in other ways. ECPs can also be used to determine local orientations and strains, but the spatial resolution is not as great as for EBSD patterns. Nevertheless the ECP technique is likely to remain a useful tool particularly for orientation determination.

Strategies for characterising bulk specimens in the SEM, which combine imaging (ECCI), orientation determination (EBSD and ECP) and phase identification (EBSD and microanalysis) in many ways parallel techniques developed for thin foils in the TEM. Despite the inferior spatial resolution, the ability to work with bulk specimens gives the SEM based diffraction techniques a powerful role to play at the mesoscale, at which many material properties are controlled.